\documentclass[aps,superscriptaddress,twocolumn]{revtex4-1}

\usepackage[pdftex]{graphicx}
\usepackage{dcolumn}
\usepackage{bm}
\usepackage{color}
\usepackage{amsmath}
\usepackage{dsfont}
\usepackage{nicefrac}
\usepackage{amssymb} 

\newcommand{\ff}[1]{{\boldsymbol #1}}
\newcommand{\ca}[1]{{\cal #1}}
\newcommand{\bi}{\begin{itemize}}
\newcommand{\ei}{\end{itemize}}
\newcommand{\be}{\begin{equation}}
\newcommand{\ee}{\end{equation}}
\newcommand{\ba}{\begin{eqnarray}}
\newcommand{\ea}{\end{eqnarray}}
\newcommand{\refe}[1]{(\ref{eq:#1})}
\newcommand{\refeq}[1]{Eq.\ (\ref{eq:#1})}
\newcommand{\labeq}[1]{\label{eq:#1}}

\newcommand{\ket}[1]{| #1 \rangle}
\newcommand{\bra}[1]{\langle #1 |}

\usepackage[pdftex,colorlinks=true,linkcolor=blue,citecolor=blue,filecolor=blue]{hyperref}

\begin{document} 
  
\title{Spin Berry curvature of the Haldane model}

\author{Simon Michel}

\affiliation{I. Institute of Theoretical Physics, Department of Physics, University of Hamburg, Notkestra{\ss}e 9-11, 22607 Hamburg, Germany}

\author{Michael Potthoff}

\affiliation{I. Institute of Theoretical Physics, Department of Physics, University of Hamburg, Notkestra{\ss}e 9-11, 22607 Hamburg, Germany}

\affiliation{The Hamburg Centre for Ultrafast Imaging, Luruper Chaussee 149, 22761 Hamburg, Germany}

\begin{abstract}
The feedback of the geometrical Berry phase, accumulated in an electron system, on the slow dynamics of classical degrees of freedom is governed by the Berry curvature. 
Here, we study local magnetic moments, modelled as classical spins, which are locally exchange coupled to the (spinful) Haldane model for a Chern insulator.
In the emergent equations of motion for the slow classical-spin dynamics there is a an additional anomalous geometrical spin torque, which originates from the corresponding {\em spin}-Berry curvature. 
Due to the explicitly broken time-reversal symmetry, this is nonzero but usually small in a condensed-matter system.
We develop the general theory and compute the spin-Berry curvature, mainly in the limit of weak exchange coupling, in various parameter regimes of the Haldane model, particularly close to a topological phase transition and for spins coupled to sites at the zigzag edge of the model in a ribbon geometry. 
The spatial structure of the spin-Berry curvature tensor, its symmetry properties, the distance dependence of its nonlocal elements and further properties are discussed in detail. 
For the case of two classical spins, the effect of the geometrical spin torque leads to an anomalous non-Hamiltonian spin dynamics. 
It is demonstrated that the magnitude of the spin-Berry curvature is decisively controlled by the size of the insulating gap, the system size and the strength of local exchange coupling.
\end{abstract} 

\maketitle 

\section{Introduction}
\label{sec:intro}

The time evolution of a quantum system with a non-degenerate and gapped ground state, steered by external time-dependent classical degrees of freedom, is governed by the adiabatic theorem \cite{Mes61,JRS07}:
If prepared at time $t=0$ in its ground state, the system state at $t>0$ is given by its instantaneous ground state for the then existing configuration of classical degrees of freedom.
Roughly, the adiabatic theorem applies, if the typical time scale $\tau$ of the classical time evolution is large compared to the inverse of the gap $\Delta E$ between the ground and the first excited state.
Besides the dynamical phase, the system accumulates a geometrical phase during the adiabatic time evolution, which cannot be gauged away in case of a cyclic motion in the classical state space \cite{Ber84,Sim83,WZ84}. 
This Berry phase and related phenomena, such as the molecular Aharonov-Bohm effect, have been studied extensively in molecular physics \cite{Mea92,Res00}, assuming that the coordinates $\ff R_{m}$ of the nuclei can be treated as classical observables.

The Berry-phase physics of the quantum system also feeds back to the dynamical state of the classical observables as has been pointed out early \cite{KI85,MSW86,Zyg87}.
In molecular systems, for example, this leads to an additional {\em geometric} force term in the classical Newtonian set of equations of motion, which has the form of a Lorentz force despite the absence of a physical magnetic field.

The geometric force plays an important role also in other contexts as, for example, in the semiclassical theory of electron dynamics in crystals, where the wave vector $\ff k$ is treated as a dynamical classical variable \cite{Res00,BMK+03}.
Its pendant on the side of the quantum system is the $\ff k$-space Berry phase, an important quantity in topological band theory
\cite{HK10,QZ11,CTSR16}.
In fact, Berry phase and geometrical force are rather general concepts that can be applied to configuration spaces of various kinds of classical degrees of freedom.

Hence, a similar situation arises for another type of quantum-classical hybrid systems as well, namely for a quantum-mechanical electron system with an exchange coupling to the local magnetic moments $\ff S_{m}$ of magnetic atoms, which are modelled as classical spins, i.e., vectors of fixed length $\ff S_{m}^{2}=1$.
The typical (picosecond) time scale for classical spin dynamics is much longer than that of the fast (femtosecond) electron quantum dynamics, such that the electron system can adiabatically follow the classical spin configuration and accumulate a finite spin-Berry phase \cite{Tat19}.
The feedback of the Berry physics on the classical system, however, is different and given by a geometrical spin torque rather than a force, and can thereby strongly affect the precessional spin dynamics \cite{SP17}. 

This spin torque is given in terms of the gauge invariant spin-Berry curvature \cite{SP17}, which must be distinguished from the 
$\ff k$-space Berry curvature and the $\ff R_{m}$-space or atomic-nuclei-Berry curvature of molecular dynamics. 
A simplified variant of the approach presented in Ref.\ \onlinecite{SP17} has been used to discuss the effect of the geometrical spin torque on the spin-wave spectrum of magnetic systems \cite{NK98,NWK+99}. 

In the recent years, the spin-Berry curvature and the geometrical spin torque have been studied more intensively.
A numerical study for an exactly solvable one-dimensional model of local magnetic moments has been carried out to quantitatively disentangle the geometrical spin torque from other contributions to the spin dynamics \cite{BN20}.
This work also emphasizes the close relation to geometric-friction effects \cite{HGT95,BR93b}, i.e., Gilbert spin damping, which can be derived within linear-response theory \cite{ON06,BNF12,SP15,SRP16a,BN19} or adiabatic-response theory \cite{CDH12}.
To go beyond the strict adiabatic limit, a generalized geometrical spin torque originating from the non-Abelian spin-Berry curvature
has been proposed recently \cite{LLP22}, in the spirit of Ref.\ \cite{WZ84}.
Effects of the geometrical spin torque in purely classical spin systems, where a single slow spin \cite{EMP20} or several slow spins \cite{MP21} interact with an exchange-coupled system of fast spins, have been studied within a formal framework largely analogous to the quantum-classical hybrid case.
This goes back to earlier work that drew attention to holonomy effects in purely classical systems \cite{Han85}. 

The main subject of this paper is to study the spin-Berry curvature for a model for local magnetic moments (classical spins) coupled to an insulator. 
The latter is described by a two-band tight-binding model for electrons hopping on a two-dimensional lattice.
With the (spinful) Haldane model on the honeycomb lattice \cite{Hal88,Ber13} we choose a prototypical model for a Chern insulator.

Our motivation for this choice is twofold: 
First, the Haldane model stands at the origin of the development topological band theory \cite{TKNN82,HK10,QZ11,CTSR16} and represents the first model for a quantum anomalous Hall insulator.
It can be experimentally realized in a setup with ultracold fermions \cite{JMD+14}.
Furthermore, there is generally a rapidly increasing interest in the study of magnetic impurities in topological insulators and at the boundaries of topological materials in particular \cite{WXX11,HKS+12,SSM+12,VPG+12,GLA13,SBK+13,EWS+14,LZLZ14,JSL+15,CTH+15,PF16,RMS+18,SKR+19}
as well as in their mutual interaction \cite{LLX+09,GCXZ09,KKB17,HKD20,YY18}. 

Second, if the Hamiltonian of the quantum system is time-reversal symmetric, there may be a finite Berry phase originating from a singularity in the classical configuration space, while the Berry curvature vanishes \cite{Ihm91}.
A finite Berry curvature ($\ff k$-, $\ff R_{m}$- or spin-Berry curvature), however, requires an explicit breaking of time-reversal symmetry. 
In case of the spin-Berry curvature in a model including an exchange coupling to the classical spins, time-reversal symmetry is already broken intrinsically, since the classical spins act like local symmetry-breaking magnetic fields. 
This effect, however, becomes irrelevant in the (physical) limit of weak exchange-interaction strength $J$.
As is shown later, the spin-Berry curvature generally vanishes in perturbation theory up to $\ca O(J^{2})$, if the quantum system itself is time-reversal symmetric.
The Haldane model is constructed such that the symmetry is broken due to a ``magnetic field'' coupling to the orbital degrees of freedom rather than to the local electron spin (note that the original model is spinless anyway), which averages to zero in a unit cell.
The mechanism by which the Haldane model leads to a nonzero Chern number can be realized in actual materials, where the spin-orbit coupling acts as a source of the orbital magnetic field \cite{RQN16,Mok18}.

The {\em spinful} Haldane model is interesting, as it allows us to study the impact of the spin-Berry curvature on the effective interaction between local magnetic moments and thus on the resulting effective spin dynamics.
The standard indirect magnetic RKKY exchange is of the Bloembergen-Rowland type \cite{BR55} in case of an insulator, and its strength depends exponentially on the insulating gap $\Delta E$, opposed to RKKY theory for metallic phases \cite{Kit68,rkky}. 
A central question is thus if the geometric spin torque is able to ``boost'' the indirect coupling between two classical spins.

To this end, we generalize the theory of Ref.\ \cite{SP17} to an arbitrary number of classical spins and derive the nonlocal geometrical spin torque that emerges in the adiabatic limit, see Sec.\ \ref{sec:theo}.
This requires to deduce explicit expressions for the spin-Berry curvature, which can be evaluated numerically, and an analysis of time-reversal and spatial symmetry transformations.
In particular, we show that the spin-Berry curvature is closely related to the spin susceptibility of the electron system.
For two classical spins, we explicitly demonstrate that the resulting adiabatic spin dynamics is anomalous and cannot be derived from an effective Hamiltonian.

On this basis, we numerically analyze the dependence of the bulk spin-Berry curvature on the positions and on the distance between two sites, at which two classical spins are exchange coupled to the electron system, see Sec.\ \ref{sec:res}.
Furthermore, we study  its dependence on the parameters of the Haldane model in the weak-$J$ regime both, for topologically trivial and nontrivial phases, and especially in the parametric vicinity of a topological phase transition. 
Invoking a phase-space argument, one can show that the spin-Berry curvature is in fact continuous across the transition. 
It diverges, however, for the Haldane model in a ribbon geometry, if the classical spins couple to sites at the zigzag edge. 

\section{Theory}
\label{sec:theo}

\subsection{Effective low-energy theory}

We consider a quantum-classical hybrid system consisting of $M$ classical spins $\ff S \equiv (\ff S_{1}, ..., \ff S_{M})$ of fixed length $S_{m} = |\ff S_{m}|=1$, the intrinsic dynamics of which are governed by a classical Hamilton function ${H}_{\text{cl}}(\ff S)$, and $N$ electrons with corresponding quantum tight-binding Hamiltonian $\hat{H}_{\text{qu}}$ constructed with the help of creation and annihilation operators $c^{\dagger}_{i \sigma}$ and $c_{i \sigma}$.
Here $i$ refers to a site of a given lattice and $\sigma = \uparrow, \downarrow$ to the electron spin projection. 
The orthonormal states $| i \sigma \rangle$ span the one-particle Hilbert space.
To develop the general theory, it is not yet necessary to further specify ${H}_{\text{cl}}(\ff S)$ and $\hat{H}_{\text{qu}}$. 
Concrete calculations will be performed for ${H}_{\text{cl}}(\ff S) = 0$ and for $\hat{H}_{\text{qu}}$ given by the Haldane model \cite{Hal88,Ber13}, but trivially generalized for spinful electrons, see Sec.\ \ref{sec:spinful}.
Spins and electrons interact via a local exchange
\be
  \hat{H}_{\text{int}}(\ff S) = J \sum_{m=1}^{M} \ff S_{m} \ff s_{i_{m}}
  \: , 
\labeq{hint}  
\ee
between the $m$th classical spin $\ff S_{m}$ and the local spin $\ff s_{i_{m}}$ of the electron system at the site $i_{m}$ given by $\ff s_i = \frac12  \sum_{\sigma\sigma'} c^{\dagger}_{i \sigma} \ff \tau_{\sigma \sigma'} c_{i \sigma'}$, where $\ff \tau = (\tau_{x},\tau_{y},\tau_{z})^{T}$ is the vector of Pauli matrices. 
With $J>0$ the coupling is antiferromagnetic.
The total Hamiltonian is
\be
\hat{H} (\ff S)
= 
\hat{H}_{\text{qu}}
+ {H}_{\text{cl}}(\ff S) 
+ \hat{H}_{\rm int}(\ff S)
\: .
\labeq{ham}
\ee

The coupled equations of motion for the spin configuration $\ff S = \ff S(t)$ and the pure $N$-electron state $|\Psi \rangle = | \Psi(t) \rangle$ are given by: 
\ba
i \partial_{t} \ket{\Psi(t)} &=& \hat{H}(\ff S)  \ket{\Psi(t)} 
\: ,
\nonumber \\
\dot {\ff S}_{m}(t) &=& \frac{\partial \langle \hat{H}(\ff S) \rangle}{\partial \ff S_m} \times \ff S_{m}(t)
\: .
\labeq{eom}
\ea
Here, $\langle \cdots \rangle$ is the expectation value with the state $\ket{\Psi(t)}$, and the dot denotes the time derivative. 
The equations can be derived from the Hamiltonian $\hat{H}(\ff S)$ or, equivalently, via the action principle, from the Lagrangian
$L = L(\ff S, \dot {\ff S} , \ket{\Psi} , \dot{\ket{\Psi}} , \bra{\Psi} , \dot{\bra{\Psi}} )$: 
\be
L = \sum_m \ff A (\ff S_m)\boldsymbol{\dot{S}}_m + \bra{\Psi}i\partial_{t}\ket{\Psi} - \bra{\Psi}
\hat{H}(\ff S) \ket{\Psi}
\: . 
\labeq{lagr}
\ee
Here,
\be
  \ff A(\ff S_{m}) = - \frac{1}{S_{m}^{2}} \frac{\ff e \times \ff S_{m}}{1+\ff e \ff S_{m} / S_{m}} \; , 
\labeq{mono}
\ee
with a fixed unit vector $\ff e$, which is usually chosen as $\ff e = \ff e_{z}$ \cite{Dir31}.
We have 
\be
\nabla \times \ff A(\ff S_{m}) = - \ff S_{m} / S_{m}^{3}
\:, 
\labeq{rot}
\ee
and thus $\ff A(\ff S_{m})$ can be interpreted as the vector potential of a unit magnetic monopole located at $\ff S_{m} = 0$. 
For the derivation of \refeq{eom} from \refeq{lagr} see the supplemental material of Ref.\ \cite{EMP20}.

The Lagrangian formulation is convenient to implement the central constraint 
\be
  |\Psi(t) \rangle = | \Psi_{0}(\ff S(t) ) \rangle
\labeq{const}  
\ee
of adiabatic spin dynamics (ASD) theory \cite{SP17,EMP20,MP21}. 
This is a holonomic constraint without explicit time dependence. 
We thereby assume that the fast electron dynamics is perfectly constrained to the manifold of instantaneous ground states $| \Psi_{0}(\ff S) \rangle$ of $\hat{H}(\ff S)$, and that the ground states are non-degenerate.
The smooth map $\ff S \mapsto | \Psi_{0}(\ff S ) \rangle$ is characterized by the spin-Berry connection:
\be
\ff C_{m}(\ff S ) = i \langle \Psi_{0}(\ff S ) | \frac{\partial}{\partial \ff S_{m}} | \Psi_{0}(\ff S ) \rangle
\: .
\labeq{conn}
\ee
Unter a local ($\ff S$-dependent) gauge transformation
\be
  | \Psi_{0}(\ff S) \rangle
  \mapsto 
  | \Psi'_{0}(\ff S) \rangle
  =
  e^{i \chi(\ff S)} 
  | \Psi_{0}(\ff S) \rangle
\ee
given by a smooth function $\chi(\ff S)$, the spin-Berry connection transforms as
\be
\ff C_{m}(\ff S ) \mapsto \ff C'_{m}(\ff S ) = \ff C_{m}(\ff S ) - \frac{\partial \chi(\ff S)}{\partial \ff S_{m}} 
\: , 
\ee
while the spin-Berry curvature 
\be
\Omega_{m\alpha, m'\alpha'} (\ff S )
=
\frac{\partial}{\partial S_{m\alpha}} C_{m'\alpha'}(\ff S )
-
\frac{\partial}{\partial S_{m'\alpha'}} C_{m\alpha}(\ff S )
\labeq{curv}
\ee
($\alpha,\alpha'=x,y,z$) is gauge invariant.

Using the constraint \refeq{const}, we can eliminate the electron degrees of freedom in the Lagrangian (\ref{eq:lagr}) so that a spin-only low-energy theory is obtained. 
The effective Lagrangian depends on the spin degrees of freedom only and is given by: 
\ba
L_{\rm eff}(\ff S, \dot{\ff S}) 
&=& 
\sum_m \ff A (\ff S_m) \dot{\ff S}_m 
+ 
\bra{\Psi_{0}(\ff S)} i \partial_{t} \ket{\Psi_{0}(\ff S)}
\nonumber \\
&-& 
\bra{\Psi_{0}(\ff S)}
\hat{H}(\ff S) 
\ket{\Psi_{0}(\ff S)}
- 
\sum_{m} \lambda_{m} \ff S_{m}^{2}
\: . 
\labeq{leff}
\ea
The last term, with Lagrange multipliers $\lambda_{m}$ takes care of the normalization condition $S_{m}=1$.
The Euler-Lagrange equations are straightforwardly derived.
We find
\ba
0 
&=& 
\frac{d}{dt} \frac{\partial L_{\rm eff}}{\partial \dot{\ff S}_{m}} - \frac{\partial L_{\rm eff}}{\partial \ff S_{m}} 
= \frac{\dot{\ff S}_{m} \times \ff S_{m}}{S_{m}^{3}} 
+ \frac{\partial \langle \hat{H}(\ff S) \rangle}{\partial \ff S_{m}}
\nonumber \\
 &-&
 \sum_{\alpha} \sum_{m'\alpha'} \Omega_{m\alpha m'\alpha'} \dot{S}_{m'\alpha'} \ff e_{\alpha}
+ 2 \lambda_{m} \ff S_{m} \: , 
\labeq{effeom0}
\ea
where $\ff e_{\alpha}$ is the $\alpha$th canonical unit vector, $\langle \cdots \rangle$ is the expectation value in the ground state $\ket{\Psi_{0}(\ff S)}$, and where we have made use of Eqs.\ (\ref{eq:rot}), (\ref{eq:conn}), (\ref{eq:curv}) and the antisymmetry of the spin-Berry curvature
\be
\Omega_{m\alpha, m'\alpha'} (\ff S) = - \Omega_{m'\alpha', m\alpha} (\ff S) 
\: .
\labeq{anti}
\ee
Scalar multiplication of \refeq{effeom0} with $\ff S_{m}$ yields an expression for $\lambda_{m}$. 
Taking the cross product with $\ff S_{m}$ from the right and using $\ff S_{m}^{2}=1$, we find the effective equation of motion
\be
\dot{\ff S}_{m} 
=
\frac{\partial \langle \hat{H}(\ff S) \rangle}{\partial \ff S_{m}} \times \ff S_{m}
+
\ff T_{m} \times \ff S_{m}
\; , 
\labeq{ano}
\ee
where the last term, $\ff T_{m} \times \ff S_{m}$, with 
\be
\ff T_{m}
=
\ff T_{m}(\ff S, \dot{\ff S})
=
\sum_{\alpha} \sum_{m'\alpha'} \Omega_{m'\alpha' m\alpha}(\ff S) \dot{S}_{m'\alpha'} \ff e_{\alpha} 
\labeq{geo}
\ee
is the geometric spin torque.
We note that for $M=1$ this reproduces the result given in Ref.\ \cite{SP17}.

The spin-Berry curvature represents the feedback of the geometrical Berry physics of the quantum system on the classical spin dynamics
in the deep adiabatic limit. 
In addition to the first term in \refeq{ano}, it gives rise to an anomalous spin torque. 
The strength of this geometrical spin torque is determined by the elements of the spin-Berry curvature tensor.

\begin{widetext}

\subsection{Spin-Berry curvature}

There are various useful general representations of the spin-Berry curvature. 
First, starting from the definition \refeq{curv} and using \refeq{conn}, we get
\be
\Omega_{m\alpha, m'\alpha'} (\ff S )
= i \left(
\frac{\partial \bra{\Psi_0}}{\partial S_{m\alpha}}
\frac{\partial \ket{\Psi_0}}{\partial S_{m'\alpha'}}
- 
\frac{\partial \bra{\Psi_0}}{\partial S_{m'\alpha'}}
\frac{\partial \ket{\Psi_0}}{\partial S_{m\alpha}}
\right)
=
- 2 \, \mbox{Im} \left(
\frac{\partial \bra{\Psi_0}}{\partial S_{m\alpha}}
\frac{\partial \ket{\Psi_0}}{\partial S_{m'\alpha'}}
\right)
\: ,
\labeq{rep1}
\ee
where $\ket{\Psi_0} = \ket{\Psi_0(\ff S )}$ has been written for short.

Second, let $\ket{\Psi_{n}(\ff S)}$ for $n>0$ be the (possibly degenerate) $n$th excited eigenstate of $\hat{H} (\ff S)$, i.e., 
$\hat{H} (\ff S) \ket{\Psi_{n}(\ff S)} = E_{n}(\ff S) \ket{\Psi_{n}(\ff S)}$. 
Differentiating with respect to $S_{m\alpha}$ yields the following identity, valid for $n\ne 0$: 
\be
   \langle \Psi_{0}(\ff S) | \nabla_{m\alpha} | \Psi_{n}(\ff S) \rangle 
   = 
   \frac{\bra{\Psi_{0}(\ff S)} \nabla_{m\alpha} \hat{H} (\ff S) \ket{\Psi_{n}(\ff S)}}{E_n - E_0} 
\; , 
\labeq{iden}
\ee
where $\nabla_{m\alpha} \equiv \partial / \partial S_{m\alpha}$.
Inserting a resolution of the unity with a local orthonormal basis of eigenstates $\{\ket{\Psi_{n}(\ff S)}\}_{n=0,1,...}$ in \refeq{rep1} and using \refeq{iden}, we find:
\be
\Omega_{m \alpha, m' \alpha'} (\ff S)
= 
- 2 \, \mbox{Im}  \sum_{n \neq 0}
\frac{\bra{\Psi_{0}(\ff S)} \nabla_{m \alpha} \hat{H}(\ff S) \ket{\Psi_{n}(\ff S)} \, \bra{\Psi_{n}(\ff S)} \nabla_{m'\alpha'} \hat{H}(\ff S) \ket{\Psi_{0}(\ff S)}}{(E_0(\ff S) - E_n(\ff S))^2} 
\: ,
\labeq{rep2prepare}
\ee
and with Eqs.\ (\ref{eq:hint}) and (\ref{eq:ham}),
\be
\Omega_{m \alpha, m' \alpha'} (\ff S)
= 
- 2 J^{2} \, \mbox{Im}  \sum_{n \neq 0}
\frac{\bra{\Psi_{0}(\ff S)} s_{i_{m} \alpha} \ket{\Psi_{n}(\ff S)} \, \bra{\Psi_{n}(\ff S)} s_{i_{m'}\alpha'} \ket{\Psi_{0}(\ff S)}}{(E_0(\ff S) - E_n(\ff S))^2} 
\: .
\labeq{rep2}
\ee
Note that a possibly present classical-spin-classical-spin coupling in $H_{\rm cl}(\ff S)$ does not contribute here.

Third, we note that the energy eigenstates for $J=0$ are trivially independent of the spin configuration $\ff S$. 
Expanding $\ket{\Psi_{n}(\ff S)} = \ket{\Psi^{(0)}_{n}} + \ca O(J)$ and $E_{n} = E^{(0)}_{n} + \ca O(J)$ and noting the explicit $J^{2}$ prefactor in \refeq{rep2}, the spin-Berry curvature takes the form
\be
\Omega_{m \alpha, m' \alpha'} (\ff S)
=
\Omega_{m \alpha, m' \alpha'} 
= 
- 2 J^{2} \, \mbox{Im}  \sum_{n \neq 0}
\frac{\bra{\Psi^{(0)}_{0}} s_{i_{m} \alpha} \ket{\Psi^{(0)}_{n}} \, \bra{\Psi^{(0)}_{n}} s_{i_{m'}\alpha'} \ket{\Psi^{(0)}_{0}}}{\big(E^{(0)}_0 - E^{(0)}_n \big)^2} 
+
\ca O(J^{3})
\labeq{rep3}
\ee 
in the weak-$J$ limit.
We see that the spin-Berry curvature becomes independent of the classical spin configuration in this case.

Fourth, in the weak-$J$ limit, there is a close relation with the retarded magnetic spin susceptibility of the unperturbed ($J=0$) electron system, which is defined as 
\be
  \chi_{i\alpha,i'\alpha'}(t) = - i \Theta(t) e^{-\eta t} \langle [ s_{i\alpha}(t) , s_{i'\alpha'}(0)] \rangle^{(0)}
  \; .
\ee
Here, $\Theta$ is the step function, $\eta$ is a positive infinitesimal, and $\langle \cdots \rangle^{(0)}$ denotes the expectation value with the ground state of the $J=0$ electron system. 
Furthermore, $s_{i\alpha}(t) = e^{i\hat{H}_{0}t} s_{i\alpha} e^{- i\hat{H}_{0}t}$ with $\hat{H}_{0} = \hat{H}|_{J=0}$.
Expanding the commutator and inserting a resolution of the unity with eigenstates of $\hat{H}_{0}$ yields the Lehmann representation via Fourier transformation:
\be
  \chi_{i\alpha,i'\alpha'}(\omega) = \int dt \, e^{i\omega t}  \chi_{i\alpha,i'\alpha'}(t)
  =
  \sum_{n\ne 0} 
  \left(
  \frac{
  \bra{\Psi^{(0)}_{0}} s_{i \alpha} \ket{\Psi^{(0)}_{n}} \,
  \bra{\Psi^{(0)}_{n}} s_{i'\alpha'} \ket{\Psi^{(0)}_{0}}
  }
  {
  \omega + i \eta - (E^{(0)}_{n}-E^{(0)}_{0})
  }
  -
  \frac{
  \bra{\Psi^{(0)}_{0}} s_{i' \alpha'} \ket{\Psi^{(0)}_{n}} \,
  \bra{\Psi^{(0)}_{n}} s_{i\alpha} \ket{\Psi^{(0)}_{0}}
  }
  {
  \omega + i \eta - (E^{(0)}_{0}-E^{(0)}_{n})
  }
  \right)
  \: .
\ee
The susceptibility at $\omega=0$ and its $\omega$-derivative at $\omega=0$ are given by: 
\be
  \chi_{i\alpha,i'\alpha'}(0)
  =
  - 2 \, \mbox{Re} 
  \sum_{n\ne 0} 
  \frac{
  \bra{\Psi^{(0)}_{0}} s_{i \alpha} \ket{\Psi^{(0)}_{n}} \,
  \bra{\Psi^{(0)}_{n}} s_{i'\alpha'} \ket{\Psi^{(0)}_{0}}
  }
  {
  E^{(0)}_{n}-E^{(0)}_{0}
  }
  \: , \;
  \frac{\partial}{\partial \omega} \chi_{i\alpha,i'\alpha'}(0)
  =
  -2i \, \mbox{Im} 
  \sum_{n\ne 0} 
  \frac{
  \bra{\Psi^{(0)}_{0}} s_{i \alpha} \ket{\Psi^{(0)}_{n}} \,
  \bra{\Psi^{(0)}_{n}} s_{i'\alpha'} \ket{\Psi^{(0)}_{0}}
  }
  {
 \big( E^{(0)}_{n}-E^{(0)}_{0} \big)^{2}
  }
  \: .
\labeq{chi}  
\ee
Here, for a gapped system, we can disregard the infinitesimal $i\eta$. 
Comparing with \refeq{rep3}, we conclude: 
\be
\Omega_{m \alpha, m' \alpha'} (\ff S)
=
\Omega_{m \alpha, m' \alpha'} 
=
- i J^{2}
\frac{\partial}{\partial \omega} \chi_{i_{m} \alpha,i'_{m}\alpha'}(\omega) \Big|_{\omega =0}
+
\ca O(J^{3})
\: .
\labeq{rep4}
\ee
We see that, in the weak-coupling limit, the spin-Berry curvature describes the linear magnetic response of the electron system due to a slow time-dependent perturbation and in fact represents the first nontrivial correction to a static perturbation.
\end{widetext}

\subsection{Time reversal and spin rotation}

Equations \refe{ano} and \refe{geo} tell us that anomalous adiabatic spin dynamics under a geometrical spin torque requires a finite spin-Berry curvature. 
In this context, it is remarkable that the spin-Berry curvature vanishes in the weak-$J$ regime, if the underlying quantum system is time-reversal symmetric. 

This is easily seen as follows: 
Let $\Theta$ be the usual anti-unitary representation of time reversal in the Fock space. 
We have $\Theta \ff s_{i_{m}} \Theta^{\dagger} = - \ff s_{i_{m}}$, and thus $\Theta \hat{H}_{\rm int} \Theta^{\dagger} = - \hat{H}_{\rm int}$. 
Obviously, the interaction term \refeq{hint}, involving classical spins that may be seen as local magnetic fields, explicitly breaks time-reversal symmetry.
However, this is irrelevant in the weak-$J$ limit, see Eqs. \refe{chi} and \refe{rep4}, since here the spin-Berry curvature is a property of  the unperturbed ($J=0$) electron system only. 
The unperturbed electron system is time-reversal symmetric if $[\hat{H}_{\text{qu}} , \Theta] = 0$. 
The case of a non-degenerate ground state, as considered here, implies that there is no Kramers degeneracy and that the total number of spin-$1/2$ electrons must be even.
Hence, the time-reversal operator squares to unity, $\Theta^{2}=+1$, and we can choose an orthonormal basis of time-reversal-symmetric eigenstates $| \Psi^{(0)}_{n} \rangle$ of $\hat{H}_{\text{qu}}$, i.e., we have $\Theta | \Psi^{(0)}_{n} \rangle = | \Psi^{(0)}_{n} \rangle$ for the ground state ($n=0$) and all excited states ($n>0$). 
Therewith, we have 
$\bra{\Psi^{(0)}_{0}} s_{i \alpha} \ket{\Psi^{(0)}_{n}} = \bra{\Theta \Psi^{(0)}_{0}} s_{i \alpha} \ket{\Theta \Psi^{(0)}_{n}}$ for the matrix element in \refeq{chi}. 
Anti-linearity of $\Theta$ implies
$\bra{\Theta \Psi^{(0)}_{0}} s_{i \alpha} \ket{\Theta \Psi^{(0)}_{n}} = \bra{\Psi^{(0)}_{0}} \Theta^{\dagger} s_{i \alpha} \Theta \ket{\Psi^{(0)}_{n}}^{\ast}$, and with $\Theta^{\dagger} s_{i \alpha} \Theta = - \ff s_{i \alpha}$ and $\Theta^{\dagger} \Theta = \ff 1$, we eventually find
$\bra{\Psi^{(0)}_{0}} s_{i \alpha} \ket{\Psi^{(0)}_{n}} = - \bra{\Psi^{(0)}_{0}} s_{i \alpha} \ket{\Psi^{(0)}_{n}}^{\ast}$. 
We conclude that time-reversal symmetry enforces that the matrix elements are purely imaginary and that
\be
  \frac{\partial}{\partial \omega} \chi_{i\alpha,i'\alpha'}(0) = 0 \: .
\ee
The spin-Berry curvature vanishes in the weak-$J$ limit for a time-reversal symmetric electron system.
As we are particularly interested in the geometrical spin torque, this motivates us to study systems with broken time-reversal symmetry already at $J=0$.
Beyond $\ca O(J^{2})$, time-reversal symmetry is broken explicitly. 

Next, we consider the usual unitary representation $U=U(\ff n,\varphi)=\exp(-i \ff s_{\rm tot} \ff n \varphi)$ of spin rotations on the Fock space, where the unit vector $\ff n$ defines the rotation axis and $\varphi$ the rotation angle.
The generators are given by the components of the total spin $\ff s_{\rm tot} = \sum_{i} \ff s_{i}$.
Invariance of $\hat{H}_{\text{qu}}$ under spin rotations, $[\hat{H}_{\text{qu}} , \ff s_{\rm tot}]=0$, implies that its eigenstates $\{\ket{\Psi_{n}}\}$ can be simultaneously chosen as eigenstates of $U$.
As $\ff s_{i_{m}}$ is a vector operator, we have $U^\dagger s_{i_{m}\alpha} U = \sum_{\beta} R_{\alpha\beta} s_{i_{m}\beta}$, where $R=R(\ff n,\varphi)$ is the defining SO(3) matrix representation. 
With this, the left matrix element in the expression for $\Omega_{m \alpha, m' \alpha'}$ [see Eqs.\ (\ref{eq:chi}) and (\ref{eq:rep4})] can be written as
$\bra{\Psi^{(0)}_{0}} U U^\dagger s_{i_{m} \alpha} U U^\dagger \ket{\Psi^{(0)}_{n}} 
= 
\sum_{\beta} R_{\alpha\beta}
e^{i \phi_{0}} \bra{\Psi^{(0)}_{0}} s_{i_{m}\beta} \ket{\Psi^{(0)}_{n}} 
e^{-i \phi_{n}}$, where the phase factors $e^{i \phi_{n}}$ are the corresponding eigenvalues of $U$.
These cancel with those obtained from the right matrix element and, hence, we find
$\Omega_{m \alpha, m' \alpha'} = \sum_{\beta\beta'} R_{\alpha\beta} \Omega_{m \beta, m' \beta'} R^{T}_{\beta'\alpha'}$. 
Irreducibility of $R$ and Schur's lemma thus imply 
$\Omega_{m \alpha, m' \alpha'} = \Omega_{mm'} \delta_{\alpha\alpha'}$. 
Furthermore, the antisymmetry condition \refeq{anti} requires that the matrix $\underline{\Omega}$ with elements $\Omega_{mm'}$ is skew-symmetric. 
In case of two impurity spins ($M=2$) this means
\be
(\Omega_{m \alpha, m' \alpha'}) 
= 
\underline{\Omega} \otimes \ff 1
=
\Omega 
\begin{pmatrix} 0 & 1\\ -1 & 0 \end{pmatrix} \otimes \ff 1 \: ,
\labeq{omega}
\ee
i.e., the spin-Berry curvature is fully determined by a single real number $\Omega$.
Note that spin-rotation symmetry requires $\Omega=0$ in the case of a single impurity spin ($M=1$), i.e., 
a finite spin-Berry curvature is obtained beyond the weak-$J$ limit only.

\subsection{RKKY interaction}

The static ($\omega=0$) magnetic susceptibility $\chi_{i\alpha,i'\alpha'}(0)$ is generally nonzero, even if $\hat{H}_{\text{qu}}$ is time-reversal symmetric and the matrix elements are purely imaginary.
This quantity just describes the linear-in-$J$ response of the electron system at time $t$.
In the adiabatic limit, the response of the electron system is in fact static: 
The local magnetic moment $\langle \ff s_{i_{m}} \rangle$ at site $i_{m}$ in the (instantaneous) ground state $| \Psi_{0}(\ff S(t))\rangle$ at time $t$, which is induced by the classical perturbations $J \ff S_{m'}(t)$ that couple to $\ff s_{i_{m'}}$, is given by:
\be
  \langle s_{i_{m}\alpha} \rangle = \sum_{m' \alpha'} \chi_{i_{m}\alpha,i_{m'}\alpha'}(0) J S_{m'\alpha'}(t) + \ca O(J^{2})
  \: .
\labeq{linres}  
\ee
Invariance of $\hat{H}_{\text{qu}}$ under SU(2) spin rotations implies that $\chi_{i_{m}\alpha,i_{m'}\alpha'}(0) = \delta_{\alpha\alpha'} \chi_{i_{m} ,i_{m'}}(0) \equiv \delta_{\alpha\alpha'} \chi_{mm'}(0)$, so that we can write $\langle \ff s_{i_{m}} \rangle = \sum_{m'} J \chi_{m,m'}(0) \ff S_{m'}(t)$ up to linear order in $J$.
With the Hellmann-Feynman theorem we have $\partial\langle \hat{H}(\ff S) \rangle  / \partial \ff S_{m} = J \langle \ff s_{i_{m}} \rangle + \partial H_{\rm cl}(\ff S) / \partial \ff S_{m}$. 
Assuming that $H_{\rm cl}\equiv 0$, for the sake of simplicity, and using Eqs.\ (\ref{eq:ano}) and (\ref{eq:linres}), we find
\be
\dot{\ff S}_{m} 
=
\sum_{m'}
J^{2} \chi_{m,m'}(0) \ff S_{m'} \times \ff S_{m}
+
\ff T_{m}(\ff S, \dot{\ff S}) \times \ff S_{m}
\; . 
\labeq{anos}
\ee
For a time-reversal-symmetric Hamiltonian, the second, geometrical term vanishes while the first is just the standard classical spin dynamics described by the effective classical Hamiltonian $H_{\rm RKKY} = \sum_{mm'} J^{\rm RKKY}_{mm'} \ff S_{m} \ff S_{m'}$ with effective RKKY \cite{rkky} exchange coupling $J^{\rm RKKY}_{mm'} =  J^{2} \chi_{mm'}(\omega=0)$.
We note, that both spin torques in \refeq{anos}, the RKKY torque and the geometric torque, come with the same prefactor $J^{2}$.

\subsection{Effective dynamics of two classical spins}
\label{sec:sdyn2}

Equation (\ref{eq:anos}) is an implicit nonlinear system of differential equations. 
It can be further simplified in the case of two classical spins.
With Eqs.\ (\ref{eq:ano}) and (\ref{eq:geo}) we get for $M=2$:
\ba
\dot{\ff S}_{1} = J^{\rm RKKY} \ff S_{2} \times \ff S_{1} - \Omega \dot{\ff S}_{2} \times \ff S_{1}
\: , 
\nonumber \\
\dot{\ff S}_{2} = J^{\rm RKKY} \ff S_{1} \times \ff S_{2} + \Omega \dot{\ff S}_{1} \times \ff S_{2}
\: .
\labeq{rkkyeom}
\ea
These equations hold in the weak-$J$ limit with $J^{\rm RKKY} = J^{2} \chi_{12}(\omega=0)=J^{2} \chi_{21}(0)$, where we can explicitly make use of \refeq{omega}.
Note that for finite $\Omega$ they are not forminvariant under exchange of the two spins $1 \leftrightarrow 2$. 
This is interesting, since any Hamilton function ${H}_{\text{cl}}(\ff S_{1},\ff S_{2})$ for identical classical spins would be symmetric, ${H}_{\text{cl}}(\ff S_{1},\ff S_{2}) = {H}_{\text{cl}}(\ff S_{2},\ff S_{1})$, and would lead to forminvariant equations of motion. 
In fact, the asymmetry of the second term $\propto \Omega$ demonstrates the anomalous, non-Hamiltonian character of the resulting spin dynamics.
A similar discussion has been given earlier \cite{MP21} in a different (purely classical) context.

For $m=1,2$ we introduce real and antisymmetric $3\times 3$ matrices 
\be
  \ca A_{m} 
  = 
  \begin{pmatrix} 
   0 & - S_{mz} & S_{my} \\
   S_{mz} & 0 & - S_{mx} \\
   - S_{my} & S_{mx} & 0 \\
  \end{pmatrix}
  \: , 
\ee
such that we can replace the cross product by a matrix-vector multiplication, $\ff S_{m} \times (...) = \ca A_{m} (...)$, and write
\ba
\dot{\ff S}_{1} &=& J^{\rm RKKY} \ff S_{2} \times \ff S_{1} + \Omega \ca A_{1} \dot{\ff S}_{2} 
\: , 
\nonumber \\
\dot{\ff S}_{2} &=& J^{\rm RKKY} \ff S_{1} \times \ff S_{2} - \Omega \ca A_{2} \dot{\ff S}_{1}
\: .
\labeq{rkkyeqn}
\ea
Using this notation, we can formally solve for $\dot {\ff S}_{1}$ and $\dot {\ff S}_{2}$, 
\be
  \begin{pmatrix} 
  \dot{\ff S}_{1} 
  \\
  \dot{\ff S}_{2} 
  \end{pmatrix}
  =
  J^{\rm RKKY} 
  \ca M^{-1}
  \begin{pmatrix} 
  \ca A_{2} & 0 \\ 
  0 & \ca A_{1}
  \end{pmatrix}  
  \begin{pmatrix} 
  \ff S_{1} \\ 
  \ff S_{2}
  \end{pmatrix}  
  \; , 
  \ee
where the $6 \times 6$ matrix $\ca M$ is defined as
\be
\ca M \equiv 
  \ff 1 - \Omega
  \begin{pmatrix} 
  0& \ca A_{1} \\ 
  - \ca A_{2} & 0
  \end{pmatrix}  
  =
  \begin{pmatrix} 
  \ff 1 & - \Omega \ca A_{1} \\ 
  \Omega \ca A_{2} & \ff 1
  \end{pmatrix}  
  \: . 
\ee
Therewith we have rewritten the equations of motion as explicit differential equations.
We note that the determinant of the block matrix is
\be
\det
\ca M
=
\det(
\ff 1 + \Omega^{2} \ca A_{2} \ca A_{1}
) \: .
\ee
A straightforward calculation yields the simple result
\be
\det
\ca M
=
\Omega^{4} (\ff S_{1} \ff S_{2})^{2} - 2 \Omega^{2} \ff S_{1} \ff S_{2} + 1
\: , 
\ee
i.e., matrix inversion is possible unless $\det \ca M =0$ or
\be
\ff S_{1} \ff S_{2} \Omega^{2} = 1
\: .
\labeq{singu}
\ee
The inversion can be done analytically:
\ba
\ca M^{-1}
=
\frac{1}{1 - \Omega^2 \ff S_1 \ff S_2}
\begin{pmatrix} 
\ff 1 - \Omega^{2} \ff S_{2} \! \otimes \! \ff S_{1} &  \Omega \ca A_{1} \\
- \Omega \ca A_{2} & \ff 1 - \Omega^{2} \ff S_{1} \! \otimes \! \ff S_{2} 
\end{pmatrix}
, 
\nonumber \\
\ea
where $\ff S_{1} \otimes \ff S_{2}$ is the outer (dyadic) product of $\ff S_{1}$ with $\ff S_{2}$.
Using the identity $\ff S_{1} \otimes \ff S_{2} \cdot \ff S_{1} \times \ff S_{2} = 0$ and combining the results, we find
\ba
\dot{\ff S}_{1} 
=
\frac{J^{\rm RKKY}}{1 - \Omega^2 \ff S_1 \ff S_2}
\left[ 
\ff S_{2} \times \ff S_{1} 
+
\Omega \ff S_{1} \times (\ff S_{1} \times \ff S_{2} )
\right] \: , 
\nonumber \\
\dot{\ff S}_{2} 
=
\frac{J^{\rm RKKY}}{1 - \Omega^2 \ff S_1 \ff S_2}
\left[ 
\ff S_{1} \times \ff S_{2} 
-
\Omega \ff S_{2} \times (\ff S_{2} \times \ff S_{1} )
\right] \: .
\nonumber \\
\labeq{seoms}
\ea
We immediately see that $|\ff S_{1}|$, $|\ff S_{2}|$, and the scalar product $\ff S_{1} \ff S_{2}$ are conserved.
The total impurity spin $\ff S_{1} + \ff S_{2}$ is not conserved for $\Omega \ne 0$.
This is compensated by a respective dynamics of the total quantum spin, see Ref.\ \cite{MP21} for a discussion in the purely classical case.

For $\Omega=0$ we recover the standard RKKY dynamics governed by the RKKY coupling constant
$J^{\rm RKKY}_{12} = J^{\rm RKKY}_{21} = J^{2} \chi_{12}(\omega=0)$.
Here, $\ff S_{1}$ and $\ff S_{2}$ precess around $\ff S_{1} + \ff S_{2} = \mbox{const}$ with frequency $\omega_{\rm L} = J_{\rm RKKY} |\ff S_{1} + \ff S_{2}|$.
The spin dynamics in that case is a Hamiltonian dynamics and derives from the effective RKKY Hamiltonian $H_{\rm RKKY} = J_{\rm RKKY} \ff S_{1} \ff S_{2}$.

For finite $\Omega$, we can distinguish between two additional effects: 
First, there is an overall renormalization of the RKKY coupling 
$J^{\rm RKKY}_{12} \mapsto J^{\rm RKKY}_{12} / (1 - \Omega^2 \ff S_1 \ff S_2)$, which depends on the initial spin configuration.
Second, there is an additional (non-Hamiltonian) coupling between the spins, the relative strength of which is given by $\Omega$.
We note that for $\Omega \to \infty$ the additional coupling cannot outweigh the renormalization effect, see \refeq{seoms}, and there is no spin dynamics at all in this limit.

Equation (\ref{eq:singu}) shows that the theory must break down for model parameters, where $\Omega = \Omega_{\rm c} \equiv 1/ (\ff S_{1} \ff S_{2})$ (recall that $\ff S_{1} \ff S_{2}$ is conserved).
For a spin-Berry curvature right at the critical value $\Omega_{\rm c}$, the dynamics of the two spins gets arbitrarily fast, such that the adiabatic theorem does not apply. 
On the other hand, \refeq{singu} or Eqs.\ (\ref{eq:seoms}) show that the strongest renormalization effect due to the geometrical spin torque occurs for model parameters, where $\Omega$ is close to $\Omega_{\rm c}$, while the exchange coupling $J$ at the same time must satisfy a condition
\be
\frac{1}{\tau} \equiv J^{2} \frac{\chi_{12}(0)}{1-\ff S_{1} \ff S_{2} \Omega^{2}} \ll \Delta E 
\labeq{criterion}
\ee
to ensure the applicability of the adiabatic theorem.
As a rule of thumb, for $M=2$ spins and for a generic value of $\ff S_{1} \ff S_{2}$ in the initial state, the value of the (with $\hbar \equiv 1$) dimensionless quantity $\Omega$ should be $\Omega = \ca O(1)$ to get a strong effect of the geometric torque.

Introducing the dimensionless time scale $t' = t / \tau$, the effective equations of motion can be rewritten in a form that is independent of all model parameters, except for the spin-Berry curvature:
\ba
\frac{d{\ff S}_{1}}{dt'}
=
\ff S_{2} \times \ff S_{1} 
+
\Omega \ff S_{1} \times (\ff S_{1} \times \ff S_{2} )
\: , 
\nonumber \\
\frac{d{\ff S}_{2}}{dt'}
=
\ff S_{1} \times \ff S_{2} 
-
\Omega \ff S_{2} \times (\ff S_{2} \times \ff S_{1} )
\: .
\labeq{seomss}
\ea
Let us emphasize once more that these equations cannot be derived from a Hamilton function $H=H(\ff S_{1}, \ff S_{2})$. 

The nonlinear system of equations (\ref{eq:seomss}) is easily solved analytically.
We define
\be
  \ff \Sigma = \ff S_{1} + \ff S_{2} + \Omega \ff S_{2} \times \ff S_{1} \: .
\ee
Using \refeq{seomss}, $d\ff \Sigma/dt = 0$ is obtained straightforwardly.
Furthermore, since $d(\ff S_{1} \ff S_{2})/dt=0$, the enclosed angle $\varphi$ is conserved.
Both, $\ff S_{1}$ and $\ff S_{2}$ undergo a precession around $\ff \Sigma$, i.e., we have $\dot{\ff S}_{1} = \ff \Sigma \times \ff S_{1}$ and $\dot{\ff S}_{2} = \ff \Sigma \times \ff S_{2}$ with frequency $\omega_{\rm prec} = \sqrt{\ff \Sigma^{2}}$. 
With $|\ff S_{1}|=|\ff S_{2}|=1$, we find
\be
  \omega_{\rm prec} = \sqrt{4\cos^{2}(\varphi/2) + \Omega^{2} \sin^{2} \varphi} \: .
\ee
For $\Omega=0$ the conventional RKKY dynamics is recovered.

\subsection{Computation of the spin-Berry curvature}

The actual calculations of the spin Berry curvature are performed for a gapped tight-binding model of non-interacting electrons with Hamiltonian
\ba
  \hat{H}_{\rm qu} 
  &=& 
  \sum_{ii' ,\sigma} t_{ii'} c_{i\sigma}^{\dagger} c_{i'\sigma}
  =
  \sum_{\ff Ir, \ff I'r' ,\sigma} t_{\ff Ir,\ff I'r'} c_{\ff Ir\sigma}^{\dagger} c_{\ff I'r'\sigma}
  \nonumber \\
  &=&
  \sum_{\ff k} t_{rr'}(\ff k) c^{\dagger}_{\ff kr\sigma} c_{\ff kr'\sigma}
  =
  \sum_{\ff k\nu} \varepsilon_{\nu}(\ff k) c^{\dagger}_{\ff k\nu\sigma} c_{\ff k\nu\sigma}
\: .
  \nonumber \\
\labeq{hop}
\ea
Here, $t_{ii'}$ is the (spin-independent) hopping matrix, and we write $i=(\ff I,r)$, where $\ff I$ are the lattice translation vectors and $r$ labels the sites within a unit cell. 
For the honeycomb lattice considered below, $r=A,B$, corresponding to A- or B-sublattice sites.
The hopping matrix is diagonalized with repect to $\ff I,\ff I'$ by a unitary transformation of the form
$c^{\dagger}_{\ff kr\sigma} = \sum_{\ff I} U^{(r)}_{\ff I\ff k} c^{\dagger}_{\ff Ir\sigma}$, such that
$t_{\ff Ir,\ff I'r'} = \sum_{\ff k} U^{(r)}_{\ff I\ff k} t_{rr'}(\ff k) U^{(r')\dagger}_{\ff k\ff I'}$.
In case of a translationally symmetric lattice with periodic boundary conditions, $\ff k$ runs over the discrete set of $L$ wave vectors in the first Brillouin zone, where $L$ is the number of unit cells, and $U_{\ff I\ff k}^{(r)} = U_{\ff I\ff k} = L^{-1/2} e^{i\ff k \cdot \ff I}$.
In a second step, the $\ff k$-dependent diagonalization of 
$t_{rr'}(\ff k) = \sum_{\nu} \overline{U}_{r\nu}(\ff k) \varepsilon_{\nu}(\ff k) \overline{U}^{\dagger}_{\nu r'}(\ff k)$ is achieved by the unitary transformation $c^{\dagger}_{\ff k\nu\sigma} = \sum_{r} \overline{U}_{r\nu}(\ff k) c^{\dagger}_{\ff kr\sigma}$ for each $\ff k$. 
Here, $\nu$ is the band index, if $\hat{H}$ is translationally symmetric. 
We have $t_{\ff Ir,\ff I'r'} = \sum_{\ff k} U^{(r)}_{\ff I\ff k} \overline{U}_{r\nu}(\ff k) \varepsilon_{\nu}(\ff k) \overline{U}^{\dagger}_{\nu r'}(\ff k) U^{(r')\dagger}_{\ff k\ff I'}$. 
With the combined transformation 
\be
S_{\ff Ir,\ff k\nu} = U^{(r)}_{\ff I\ff k} \overline{U}_{r\nu}(\ff k) \: ,
\labeq{comb}
\ee 
this reads as
$t_{\ff Ir,\ff I'r'} = \sum_{\ff k\nu} S_{\ff Ir,\ff k\nu} \varepsilon_{\nu}(\ff k) S^{\dagger}_{\ff k\nu,\ff I'r'}$.
\begin{widetext}

Evaluating the matrix elements in \refeq{rep2} for $N$-electron Slater determinants $\ket{\Psi^{(0)}_{n}} = \prod_{k\nu\sigma} c_{k\nu\sigma}^{\dagger} |\mbox{vac} \rangle$ is straightforward.
We find
\be
\Omega_{m \alpha, m' \alpha'} 
= 
- J^{2} \delta_{\alpha\alpha' }\, \mbox{Im}  \sum_{\ff k\nu}^{\rm occ.}\sum_{\ff k'\nu'}^{\rm unocc.}
\sum_{rr'}
\frac{
S^{\dagger}_{\ff k\nu,\ff Ir}
S_{\ff Ir,\ff k'\nu'} 
S^{\dagger}_{\ff k'\nu',\ff I'r'}
S_{\ff I'r',\ff k\nu} 
}
{
( \varepsilon_{\nu'}(\ff k') - \varepsilon_{\nu}(\ff k) )^{2}
}
\: ,
\labeq{rep5}
\ee
where $(\ff I,r) = i_{m}$ and $(\ff I',r')=i_{m'}$.
This representation applies to the translationally symmetric model $\hat{H} = \hat{H}_{\rm qu}$ for $J=0$ and can be used for the case, where $J$ is treated perturbatively.
Beyond the weak-$J$ regime, it applies for each spin configuration $\ff S$ in $\hat{H} = \hat{H}_{\rm qu} + \hat{H}_{\rm int}(\ff S)$.
But as translational symmetry is broken in this case, $S_{\ff Ir,\ff k\nu}$ does no longer contain the Fourier factor $L^{-1/2} e^{i\ff k\ff I}$ and must be computed numerically and explicitly include the $\alpha$ index.

\end{widetext}

\subsection{Spinful Haldane model}
\label{sec:spinful}

For generic parameters the half-filled Haldane model \cite{Hal88} is a prototypical Chern insulator with broken time-reversal symmetry. 
Numerical calculations have been performed for the spinful Haldane model at half-filling, which consists of two identical copies of the Haldane model for electrons with spin projection $\sigma=\uparrow$ and $\sigma=\downarrow$, respectively. 
The spinful Haldane model is trivially invariant under SU(2) spin rotations. 
Rotation symmetry is broken in the full model \refeq{ham}, where the two copies are coupled by the interaction term \refeq{hint}.

\begin{figure}[t]
\includegraphics[width=0.65\columnwidth]{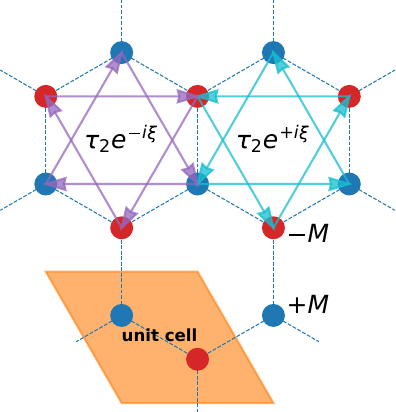}
\caption{
Haldane model on the honeycomb lattice. A sites indicated by blue dots with on-site potential $+M$. B sites: red dots, potential $-M$.
Nearest-neighbor hopping with amplitude $\tau_{1}$: dashed lines. Next-nearest-neighbor hopping $\tau_{2}$ with additional Peierls factor $e^{-i\xi}$ ($e^{i\xi}$) for hopping in clockwise (counterclockwise) direction: purple (light-blue) arrows. 
}
\label{fig:hal}
\end{figure}

The corresponding tight-binding quantum Hamiltonian in \refeq{ham} is given by
\ba
\hat{H}_{\rm qu} 
&=& 
M \sum_{i\sigma} z_{i} c_{i\sigma}^{\dagger} c_{i\sigma}
- 
\tau_{1}\!\!
\sum_{\langle ii' \rangle, \sigma} c_{i\sigma}^{\dagger} c_{i'\sigma}
\nonumber \\
&-& 
\tau_{2} \!\!
\sum_{\langle\langle ii' \rangle\rangle, \sigma} 
e^{i\xi_{ii'}}
c_{i\sigma}^{\dagger} c_{i'\sigma}
\: .
\labeq{haldane}
\ea
Here, $i$, $i'$ run over the sites of the two-dimensional bipartite honeycomb lattice, see Fig.\ \ref{fig:hal}.
$M$ is the strength of a staggered on-site potential, where the sign factor $z_{i}=+1$ for a site $i$ in the A sublattice ($r=A$) and $z_{i}=-1$ for $i$ in the B sublattice ($r=B$).
For $M\ne 0$ the potential term breaks particle-hole symmetry. 
The amplitude for hopping between nearest-neighbor sites on the lattice, $\langle ii' \rangle$, is denoted by $\tau_{1}$.
We set $\tau_{1}=1$ to fix the energy scale (and with $\hbar=1$) also the time scale.
The next-nearest-neighbor hopping amplitude $e^{i \xi_{ii'}} \tau_{2}$ with real $\tau_{2}$ includes a phase factor with $\xi_{ii'} = -\xi$ for hopping from $i'$ to $i$ in clockwise direction (see purple lines in Fig.\ \ref{fig:hal}), and with $\xi_{ii'} = \xi$ in counterclockwise direction (light blue).
This ensures that the total flux of the corresponding orbital magnetic field through a unit cell vanishes.
Time-reversal symmetry is broken for $\tau_{2}\ne 0$ and $\xi \ne 0, \pm \pi$.

\begin{figure}[t]
\includegraphics[width=0.9\columnwidth]{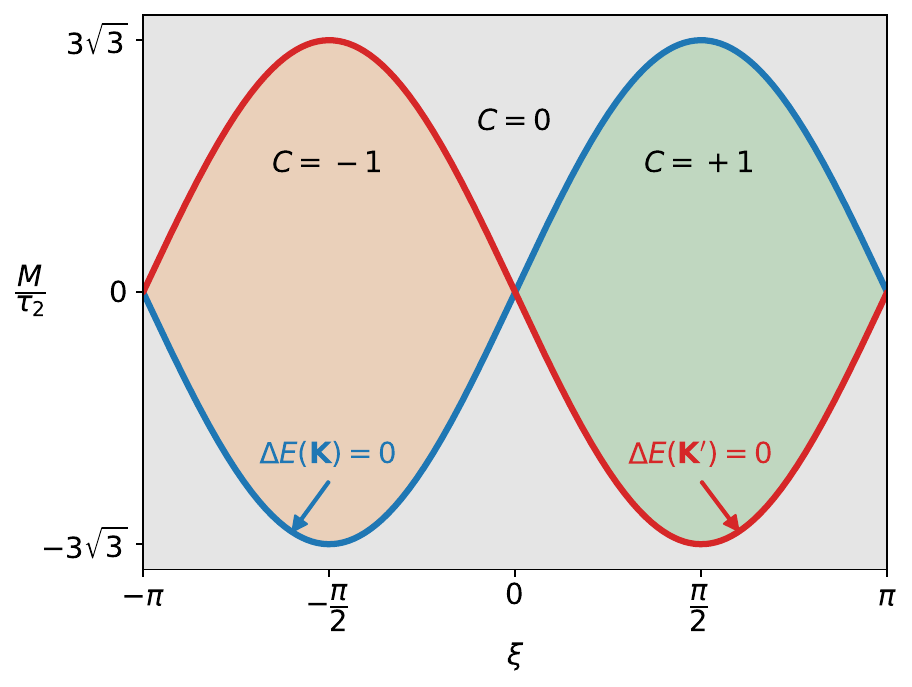}
\caption{
Topological phase diagram of the Haldane model in the $\xi$-$M/\tau_{2}$ plane.
$C$ denotes the first Chern number.
The nearest-neighbor hopping $\tau_1 = 1$ sets the energy scale. 
}
\label{fig:tpd}
\end{figure}

The band structure of the spinful Haldane model consists of two bands with trivial twofold spin degeneracy each.
The band structure is gapped for all $M$ and all $\xi$, if $|\tau_{2}/\tau_{1}| < 1/3$, except for parameter values satisfying the condition 
\cite{Hal88} $| M/\tau_{2} | = 3 \sqrt{3} | \sin \xi |$, or, in terms of the band gap given by
\be
   \Delta E = | M - 3 \sqrt{3} \tau_{2}  \sin \xi  | \stackrel{!}{=} 0 \: ,
\labeq{tpt}   
\ee
for $M, \tau_{2} >0$ and $0<\xi < \pi$.
In the $M/\tau_{2}$ vs.\ $\xi$ phase diagram, see Fig.\ \ref{fig:tpd}, this condition defines lines of topological phase transitions (see red and blue lines in the figure).
At a phase transition, there is a band closure in the first Brillouin zone, either at $K=2\pi (1/3\sqrt{3},1/3)$, when $M=-3\sqrt3 \, \tau_2\sin(\xi)$, or $K'=2\pi (2/3\sqrt{3},0)$, when $M=3\sqrt3 \, \tau_2\sin(\xi)$ \cite{Ber13}.

In the Altland-Zirnbauer classification \cite{AZ97}, the model belongs to symmetry class A.
For parameters inside the boundary in Fig.\ \ref{fig:tpd}, i.e., in the orange or green regions for $| M/\tau_{2} | < 3 \sqrt{3} | \sin \xi |$, the model represents a Chern insulator with finite Chern number $C=-1$ ($\xi < 0$) and $C=+1$ ($\xi >0$).
Outside the boundary, we have $C=0$, and the system is a topologically trivial band insulator. 

The Haldane model $\hat{H}_{\text{qu}}$ is invariant under SU(2) spin rotations, i.e., we can make use of \refeq{omega}. 
Furthermore, it is invariant under the discrete $C_{3}$ rotations of the lattice around a fixed site, and thus $\Omega$ is $C_{3}$ invariant. 
Time-reversal symmetry, on the other hand, is broken explicitly.
Under time reversal, we have
$\bra{\Psi^{(0)}_{0}} s_{i \alpha} \ket{\Psi^{(0)}_{n}} \mapsto - \bra{\Psi^{(0)}_{0}} s_{i \alpha} \ket{\Psi^{(0)}_{n}}^{\ast}$. 
With \refeq{rep3}, this implies $\Omega_{m \alpha, m' \alpha'} \mapsto \Omega_{m' \alpha',m \alpha}$, and using \refeq{omega} we find $\Omega \mapsto -\Omega$.
Under a reflection at a mirror symmetry axis of the hexagonal lattice, the Hamiltonian transforms in the same way as under time reversal, $H \mapsto \Theta H \Theta^{\dagger}$. 
The first two terms on the right-hand side of \refeq{haldane} are invariant. 
Reflection of the hopping term for two next-nearest-neighbor sites $i, j$ is represented by complex conjugation, since $\xi_{ij}=-\xi_{ji}$. 
Hence, we have
\be
\Omega \mapsto -\Omega \quad \mbox{under time reversal or reflection}
\labeq{sym}
\: .
\ee
The Hamiltonian and thus $\Omega$ is invariant under the combined transformation. 

\subsection{Spin-Chern number}

Let us emphasize that the different topological phases of the Haldane model are characterized by the first Chern number $C$, which is obtained by integrating the conventional $\ff k$-space Berry curvature $F(\ff k)$ over the entire BZ, i.e., the torus $T^{2}$. 
The {\em spin}-Berry curvature is a different concept, motivated by the geometrical spin torque contribution to the adiabatic impurity-spin dynamics. 

Still it can be employed to define a topological invariant: 
The Hamiltonian of the quantum system smoothly depends on the parameters $\ff S = (\ff S_{1}, ..., \ff S_{M}) \in \ca S$, where $\ca S$ is the Cartesian product of two-spheres 
(recall $|\ff S_{r}|=1$). 
The parameter manifold $\ca S$ is closed, i.e., has no boundary, and is $2M$-dimensional.
Hence, according to the general theory (see Ref.\ \cite{Nak98}, for example), the $M$th spin-Chern number $C^{(S)}_{M}$ is given by 
\be
  C^{(S)}_{M} = \frac{i^{M}}{(2\pi)^{M}} \frac{1}{M!} \oint_{S} \mbox{tr} \, \omega^{M} \: ,
\ee
where $\omega = dA + A^{2}$ is the spin-Berry-curvature two-form derived from the one-form $A$, the spin-Berry connection.
If the ground state of the Hamiltonian is nondegenerate on the entire manifold $\ca S$, the spin-Chern number $C^{(S)}_{M}$ is well-defined and quantized.

\begin{figure*}[t]
\includegraphics[width=1.99\columnwidth]{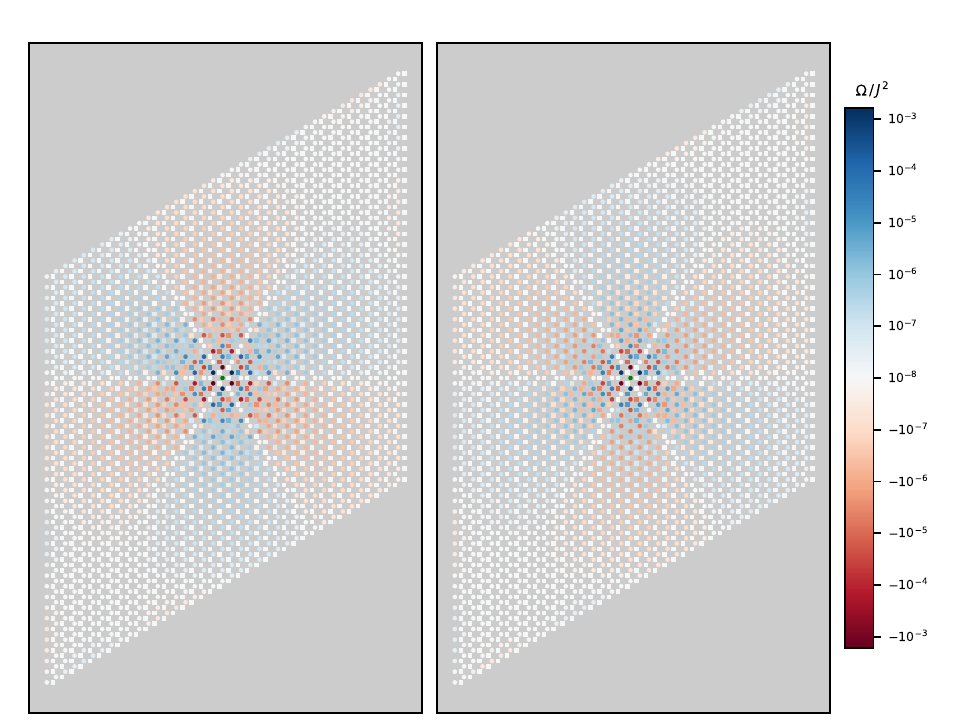}
\caption{
Nonzero element $\Omega$ of the spin-Berry curvature $\Omega_{m \alpha, m' \alpha'}$ [see Eq.\ \ref{eq:omega}] for various positions $i_{m}$ and $i_{m'}$ of the two impurity sites, to which the classical spins are coupled. 
The first spin resides at the central site $i_{1}$ (see green dot), and the second one is varied over the remaining $2L-1$ sites of the hexagonal lattice.
Calculations for a lattice with $L=39 \times 39$ unit cells and periodic boundary conditions.
Model parameters: $\tau_2 = 0.1$, $\xi = \pi / 4$. 
$\tau_1 = 1$ sets the energy scale. 
{\em Left:} $M = 0.8 M_{\rm crit}$, such that the system is in a topologically nontrivial phase.
{\em Right:} $M = 1.2 M_{\rm crit}$, topologically trivial phase.
Results for the weak-$J$ regime. 
The color code indicates the value of $\Omega/J^{2}$.
}  
\label{fig:pos}
\end{figure*}

The computation of $C^{(S)}_{M}$ is a highly nontrivial task. 
Here we note that $C^{(S)}_{M} = 0$ in the weak-$J$ limit, on which we concentrate in the present study. 
This is easily verified, since for $J=0$ the spin-Berry curvature is {\em independent} of $\ff S$ (see \refeq{rep3}) and hence $C^{(S)}_{M}(J=0)=0$.
Quantization of the spin-Chern number and continuity with respect $J$ then imply $C^{(S)}_{M}(J) = 0 + \ca O(J^{2})=0$. 

On the other hand, in the strong-coupling limit, the interaction term, \refeq{hint}, will dominate the physics.
In the case of a single classical spin, the first Chern number of the ``atomic'' model, $\hat{H} = \hat{H}_{\text{int}}(\ff S) = J \ff S \ff s$ is $C^{(S)}_{1}=\pm 1$ \cite{SP17,Ber13}.
Generally, it is thus plausible that there is a nonzero spin-Chern number $C^{(S)}_{M}$ for $J \to \infty$.
Hence, as a function of $J$ we expect a topological phase transition and thus an accompanying gap closure, which, due to the vanishing energy denominator in \refeq{rep2}, may have a substantial effect on the magnitude of the spin-Berry curvature close to transition.
Interestingly, our data presented below (see Sec.\ \ref{sec:strong}) indeed demonstrate a qualitative difference between the weak and the strong-coupling limit but do not hint towards singular behavior of the spin-Berry curvature.
Further research along this line is in progress.

\section{Numerical Results}
\label{sec:res}

The spin-Berry curvature of the Haldane model can be computed numerically using the representation \refeq{rep5}, if the exchange coupling $J$ is weak.
As mentioned above in the discussion of \refeq{rep5}, a slightly generalized formula applies to the general, non-perturbative case. 
As translational symmetry is broken in this case, however, the accessible system sizes are considerably smaller.
We will, therefore, start with a discussion of results, obtained for a bulk system in the weak-$J$ limit.

\subsection{Spatial structure of the spin-Berry curvature}

The $m, m'$ element of the spin-Berry curvature tensor $\Omega_{m \alpha, m' \alpha'} = \Omega_{mm'} \delta_{\alpha\alpha'}$ describes the strength of the mutual geometric spin torque of two classical spins $\ff S_{m}$ and $\ff S_{m'}$, locally exchange coupled to local spins of the electron system at two sites $i_{m}$ and $i_{m'}$ of the lattice. 
Since $\Omega_{mm'}$ is antisymmetric, it is sufficient to specify a single real number $\Omega$ for each pair of lattice sites, see \refeq{omega}.
We consider a translationally invariant system with periodic boundary conditions, fix the position of one site, and compute the spin-Berry curvature $\Omega$ as a function of the position of the second site, using \refeq{rep5}.

In Fig.\ \ref{fig:pos}, the fixed site in the center of the plot is marked with green color, and the $J$-independent quantity $\Omega / J^{2}$ is given by the color code on each of the remaining lattice sites. 
Sites of the A sublattice are indicated by circles, those of the B sublattice sites by squares. 
A large lattice is chosen with $L = 39 \times 39$ unit cells, each consisting of two sites, i.e., with a total number of $2L=3042$ sites. 
Calculations have been done for parameters $\tau_2 = 0.1$ and $\xi = \pi / 4$, where the system is in the topologically nontrivial and the trivial phase, with $M = 0.8 M_{\rm crit}$ (left plot) and $M = 1.2 M_{\rm crit}$ (right plot), respectively. 
For this parameter set we have $M_{\rm crit} \approx 0.37$. 
Note that the parameters are chosen such that the gap size $\Delta E$ is the same for both, the topologically nontrivial (left) and the trivial phase (right). 

\begin{figure}[t]
\includegraphics[width=0.99\columnwidth]{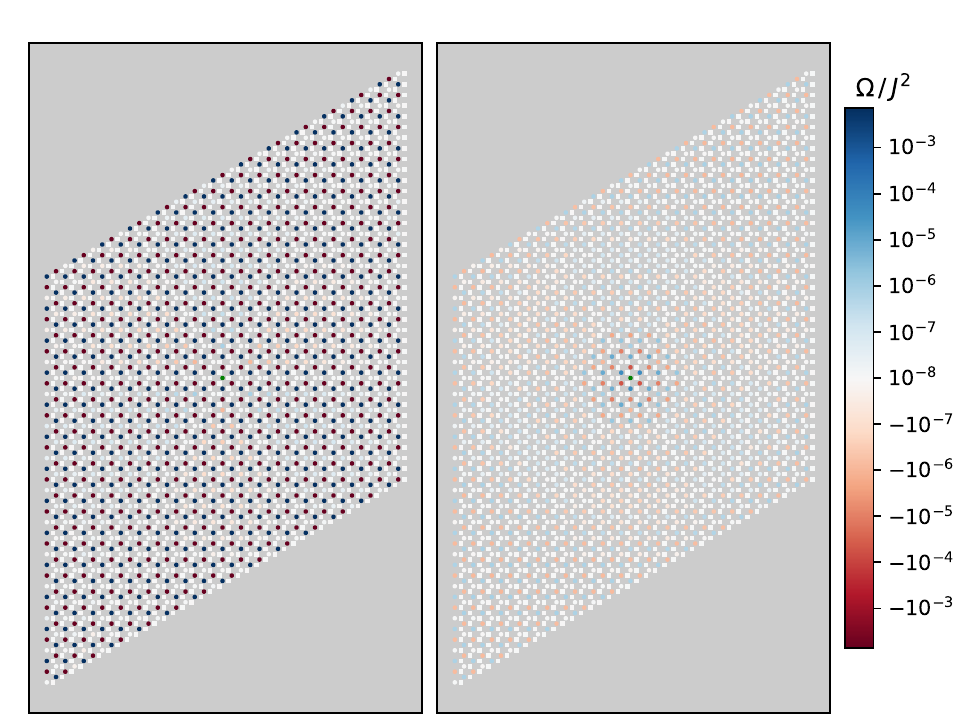}
\caption{
The same as Fig.\ \ref{fig:pos} but for $\tau_2 = 0.001$.
{\em Left:} topologically nontrivial phase at $M = 0.8 M_{\rm crit}$.
{\em Right:} trivial phase, $M = 1.2 M_{\rm crit}$.
}  
\label{fig:pos1}
\end{figure}

One can nicely see the invariance of $\Omega$ under discrete $2\pi/3$ rotations of the lattice around the fixed site.
This set $C_{3}$ of spatial rotations forms in fact a symmetry group of the Haldane model. 
Furthermore, consistent with \refeq{omega}, $\Omega$ changes sign if $i_{m}$ and $i_{m'}$, i.e., $m$ and $m'$ are exchanged.
In the figure, where site $i_{m}$ is kept fixed, this is seen be comparing $\Omega$ for A-sublattice sites $i_{m'}$ at opposite positions $\ff I_{m}$ and $-\ff I_{m'}$.
Finally, consistent with \refeq{sym}, $\Omega$ changes its sign under a reflection at the horizontal axis through the central site $i_{m}$ and at the axes rotated by $2 \pi /3$ and $-2\pi/3$ against the horizontal. 
This also implies that directly on these mirror axes $\Omega$ vanishes.

Concerning the distance dependence, we can distinguish between two different ranges. 
For small distances, up to about 3 unit cells, $\Omega$ has an oscillatory distance dependence. 
In this {\em close range}, the spatial structure is rather complicated generally. 
On the other hand, in the {\em far range} $\Omega$ does not change sign and monotonically decreases with increasing distance between $i_{m}$ and $i_{m'}$ along any spatial direction. 

We recall that, in the weak-$J$ limit, the spin-Berry curvature $\Omega$ is purely a property of the spinful Haldane model. 
However, it is not directly related to its band topology, as the weight factors in \refeq{rep5} are constructed from the same Bloch states but in a different way as compared to the $\ff k$-space Berry curvature. 
Nevertheless, the sign structure of $\Omega$ in the far range is quite universal, and it is different for the topologically nontrivial and the trivial phase, as can be seen in Fig.\ \ref{fig:pos}, although the gap $\Delta E$ is the same.
This implies that, to some degree, the spin dynamics is sensitive to the respective topological phase.
In this sense, the spin-Berry curvature can be seen as a marker for the topological properties of the model.

Differences between the two phases become more pronounced in the case of a smaller gap $\Delta E$.
Fig.\ \ref{fig:pos1} displays results for a next-nearest-neighbor hopping amplitude $\tau_{2}$, which is smaller by two orders of magnitude as compared to Fig.\ \ref{fig:pos}.
Choosing $M = 0.8 M_{\rm crit}$ and $M = 1.2 M_{\rm crit}$, respectively, as in Fig.\ \ref{fig:pos}, a small $\tau_{2}$ leads an overall small gap $\Delta E$ in the entire phase diagram, as can be inferred from \refeq{tpt}.
Again, the gap $\Delta E$ is the same for both, the topologically nontrivial and the trivial phase.

But now, for small $\tau_{2}$, the spatial structure of $\Omega$ is largely different for both phases.
In the nontrivial phase (left plot), the close range now spreads over the entire (finite) system, and there is hardly any visible decrease of $\Omega$ with increasing distance. 
In the trivial phase (right), on the contrary, the close range is limited to distances of a few unit cells only, while $\Omega$ is clearly reduced in size in the far range. 
For still larger distances from the central site, $\Omega$ increases again. 
This is due to finite-size effects, which are more pronounced for small $\Delta E$. 
Since, as compared to Fig.\ \ref{fig:pos}, the gap is smaller by two orders of magnitude, this is not unexpected.

\subsection{Finite systems in the small $\tau_{2}$ regime}

Interestingly, the absolute value of $\Omega$ in Fig.\ \ref{fig:pos1} is much larger in the nontrival case, where the far range extends over the whole lattice.
Quite generally, finite systems (with periodic boundary conditions) in the regime of small next-nearest-neighbor hopping $\tau_{2}$ are interesting 
as here the spin-Berry curvature can be extremely large. 

\begin{figure}[b]
\includegraphics[width=0.95\columnwidth]{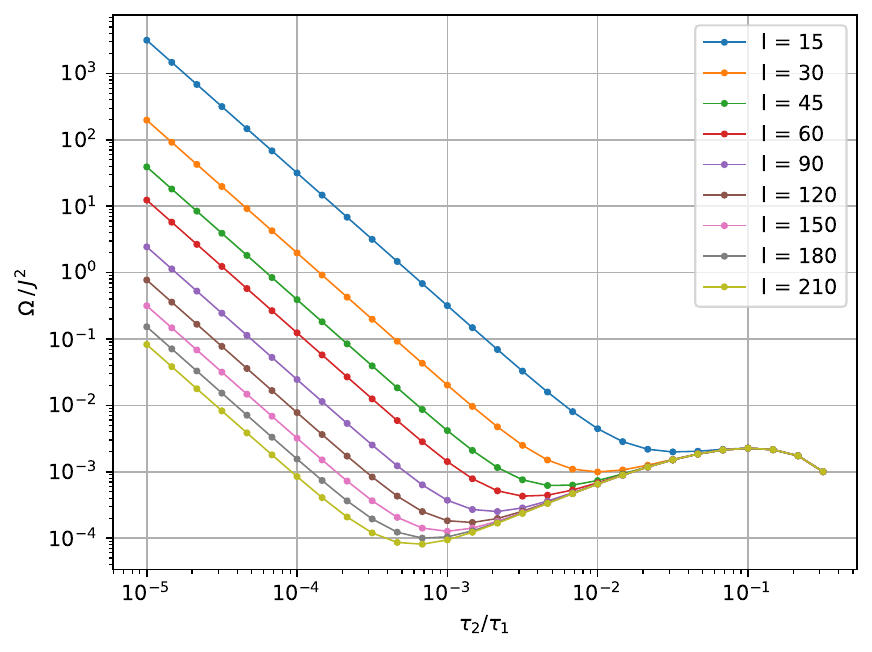}
\caption{
$\Omega$ for next-nearest-neighbor sites as function of $\tau_{2}$ for $\xi = \pi/4$ and $M=0$. 
Note the log-log scale.
Calculations for finite systems with periodic boundary conditions of different size $L = l \times l$ as indicated. 
}
\label{fig:pdtau2}
\end{figure}

This is seen in Fig.\ \ref{fig:pdtau2}, where $\Omega$ is plotted as a function of $\tau_{2}$.  
Here, we have set $M=0$, such that the model is topologically nontrivial for all values of $\tau_{2}$. 
Exactly at $\tau_{2}=0$, however, the model is a time-reversal symmetric semimetal, such that the Chern number is zero.
Hence, in the thermodynamical limit $L\to \infty$, there is a phase transition at $\tau_{2}=0$.

From the log-log plot in Fig.\ \ref{fig:pdtau2} we can infer that $\Omega \propto 1/\tau_{2}^{2}$ for $\tau_{2} \to 0$.
This has interesting consequences, as has already been discussed above in Sec.\ \ref{sec:sdyn2}: 
Namely, for $\tau_{2} \to 0$, the spin dynamics is entirely dominated by the anomalous contribution from the geometrical spin torque.
The adiabatic spin dynamics slows down, and the system in this limiting case ultimately shows no dynamics at all. 
At intermediate values for $\tau_{2}$, however, depending on the value for $J$, the value of $\Omega$ can be of order one. 
This is exactly the range, where dynamic effects of the geometrical spin torque are most pronounced, as argued in Sec.\ \ref{sec:sdyn2}.

Fig.\ \ref{fig:pdtau2} also shows that the $1/\tau_{2}^{2}$ behavior is realized for strictly finite systems only. 
Comparing the results for different $L$ with linear extension $l$ up to $l=210$ (88200 sites), we see that rather a linear dependence $\Omega(\tau_{2}) \propto \tau_{2}$ is obtained in the thermodynamical limit.
Note that this is just the $\tau_{2}$ dependence that must be expected, when expanding $\Omega(\tau_{2})$ around $\tau_{2}=0$, where the model is time-reversal symmetric and thus $\Omega=0$, up to linear order in $\tau_{2}$.
We conclude that in the thermodynamical limit $\Omega$ is continuous at the phase transition, when this is steered via $\tau_{2}$, while $\Omega$ diverges as $1/\tau_{2}^{2}$ for any finite system.

The underlying mechanism is the following:
For small systems, where the $\ff k$-space is strongly discretized, the relative contribution to $\Omega$ from regions in the Brillouin zone close to $K$ or $K'$ is comparatively large, so that this becomes the dominating contribution to $\Omega$, for model parameters close to the transition.
For larger systems, however, the relative contribution diminishes, as can be seen in Fig.\ \ref{fig:pdtau2}.
If, for a finite system, $\tau_{2}$ is sufficiently small, only the contributions from $K$ or from $K'$ are relevant in the double sum over $\ff k$ and $\ff k'$ in \refeq{rep5}.
In this case, the $\tau_{2}$ dependence of the spin-Berry curvature is essentially given via the energy denominator in \refeq{rep5} only, and since $\Delta E \propto \tau_{2}$, see \refeq{tpt}, we have the scaling $\Omega \propto 1/\tau_{2}^{2}$.

The different scalings $\Omega \propto \tau_{2}$ and $\Omega \propto 1/\tau_{2}^{2}$, distinguish between asymptotic behavior in the thermodynamic limit and for a finite-size system. 
Fig.\ \ref{fig:pos1} (left) for the nontrivial phase is in fact representative for a system, where finite-size effects dominate, 
while in the case of Fig.\ \ref{fig:pos} the system size is sufficiently large to reflect the spin-Berry curvature in the thermodynamic limit (for not too large distances).

\subsection{Distance dependence}

\begin{figure}[t]
\includegraphics[width=0.9\columnwidth]{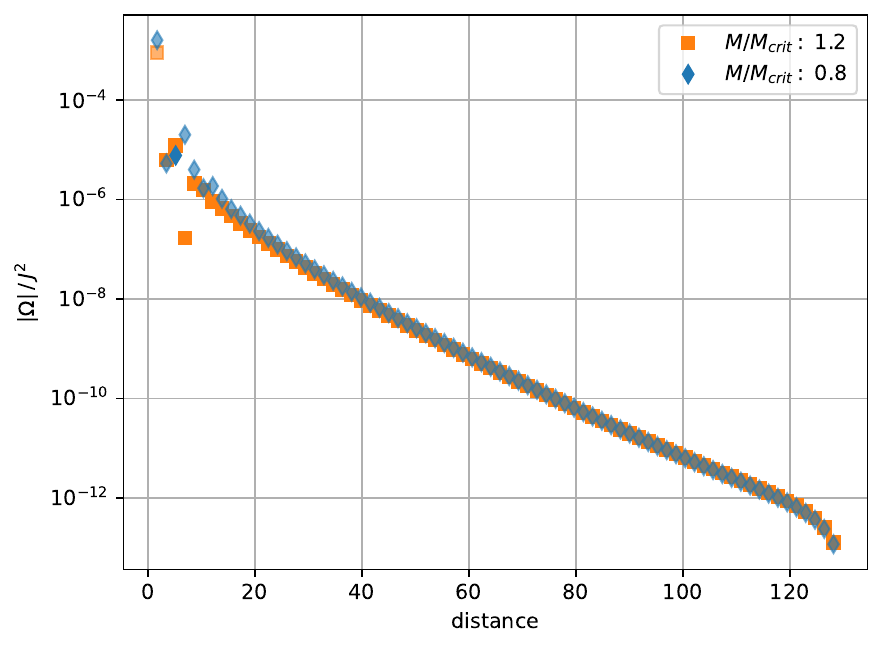}
\caption{
$\Omega / J^{2}$ as a function of the distance $d$ between the two impurity sites for the topologically nontrivial 
($M = 0.8 M_{\rm crit}$, blue diamonds) and the trivial phase ($M = 1.2 M_{\rm crit}$, orange squares) along a path including A-sublattice sites, starting from the central (green) site in Fig.\ \ref{fig:pos} in vertical direction.
Filled symbols: positive sign of $\Omega / J^{2}$. 
Hollow symbols: negative sign.
Further parameters as in Fig.\ \ref{fig:pos}, but for a larger lattice with $L =150 \times 150$ unit cells (periodic boundary conditions).
}  
\label{fig:dberry}
\end{figure}

For an analysis of the distance dependence of $\Omega$, we revert to the same parameters $\tau_{2}=0.1$ and $\xi = \pi / 4$, as underlying Fig.\ \ref{fig:pos}, but choose a larger system with $L=150 \times 150$ unit cells.
Fig.\ \ref{fig:dberry} shows the dependence of the spin-Berry curvature $\Omega$ on the distance $d$ between the two impurity sites. 
$d$ is defined as the Euclidean distance between the sites $i_{m}$ and $i_{m'}$ on the hexagonal lattice in units of the nearest-neighbor distance.
The distance of a site to all of its six next-nearest neighbors, for example, is given by $d=\sqrt3$.
For distances $40 \lesssim d \lesssim 100$, we find a nearly linear dependence of $\ln \Omega$ on $d$, i.e., $\Omega \propto \exp(-d/\lambda)$ with $\lambda>0$, 
while for too large distances $d \gtrsim 100$, the linear trend is disturbed by finite-size effects.
Note that for the larger distances only the sign of $\Omega$ depends on $M$ but not its absolute value, if the gap is the same (as is the case for $M=1.2M_{\rm crit}$ and $M=0.8M_{\rm crit}$).

\begin{figure}[b]
\includegraphics[width=0.9\columnwidth]{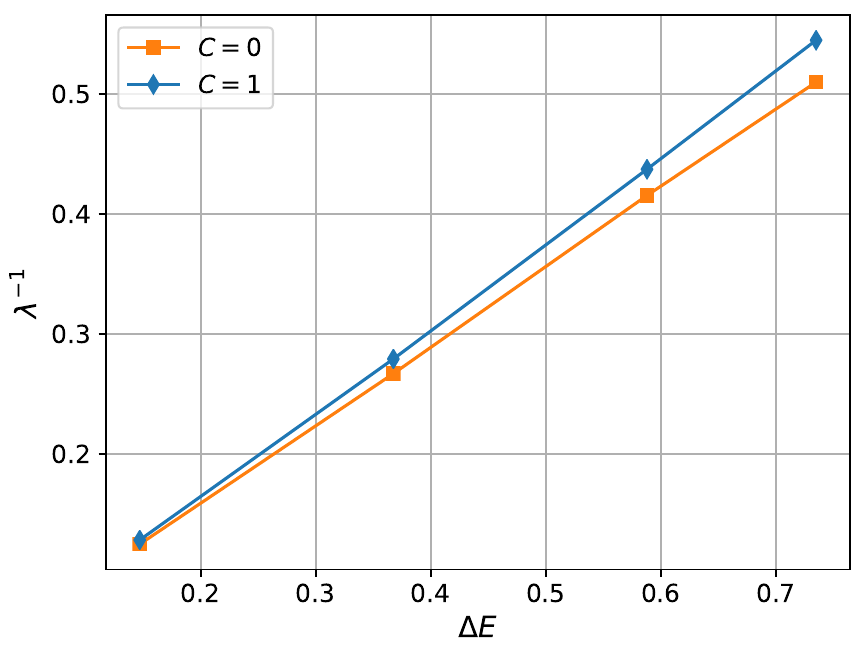}
\caption{
$1/\lambda$ as a function of the gap $\Delta E$, as obtained from calculations for systems with $L=l \times l$ with $l=150$ unit cells for different $M$.
Calculations for the nontrivial and the trivial phase with the same gap size.
See text for discussion.
}  
\label{fig:fit}
\end{figure}

\begin{figure}[t]
\includegraphics[width=0.9\columnwidth]{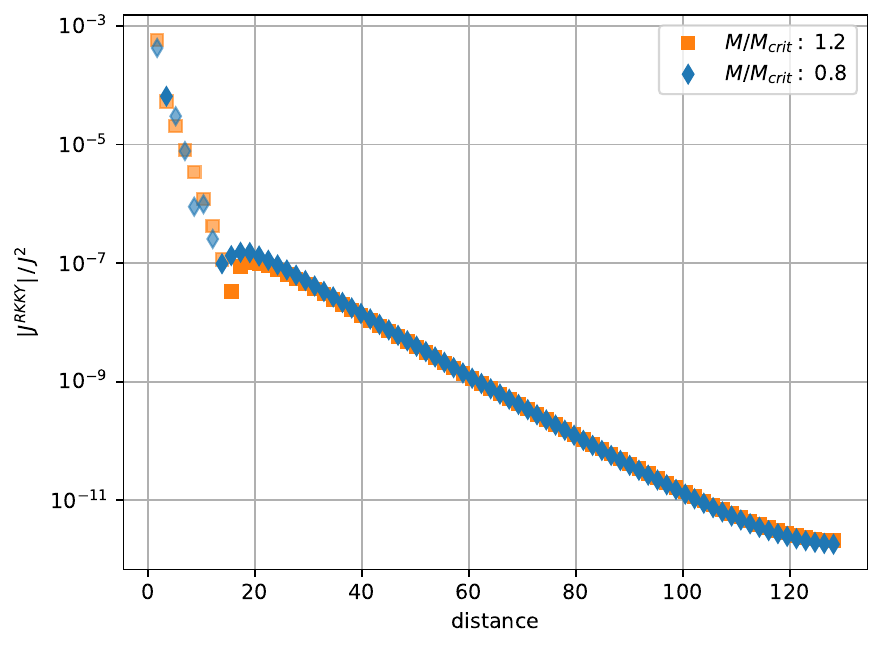}
\caption{
The same as Fig.\ \ref{fig:dberry} but for $J_{\rm RKKY} / J^{2}$ as a function of the distance $d$.
Filled symbols: positive sign of $J_{\rm RKKY} / J^{2}$. 
Hollow symbols: negative sign.
}  
\label{fig:drkky}
\end{figure}

Performing calculations for different $M$ to vary the gap $\Delta E$, we can extract the $\Delta E$ dependence of the slope $-1/\lambda$.
This is plotted in Fig.\ \ref{fig:fit}. 
We find a nearly linear dependence $1/\lambda \propto \Delta E$. 
This means that the spin-Berry curvature has an exponential $d$ dependence for large $d$, which is controlled by the bulk band gap $\Delta E$:
$\Omega \propto \exp(- \Delta E \, d)$.

This behavior is reminiscent of the exponential decay of the RKKY exchange interaction with $d$ for insulating systems \cite{BR55}. 
Fig.\ \ref{fig:drkky} gives an example. 
Here, $J_{\rm RKKY} / J^{2}$ is plotted as function of $d$ for the same model parameters as in Fig.\ \ref{fig:dberry}, and exponential behavior is found for the RKKY coupling in the same range $40 \lesssim d \lesssim 100$.

The range of distances with an exponential dependence of $\Omega$ or $J_{\rm RKKY}$ exactly corresponds to the far range seen in Fig.\ \ref{fig:pos}, where there is a comparatively smooth dependence of $\Omega$ on the position of the second impurity spin.
In the close range, for $d \lesssim 20$, the distance dependence of $\Omega$ is less regular, and there are sign changes of $\Omega$ in addition. 
This is somewhat reminiscent of the oscillatory distance dependence of the RKKY exchange for a metal or semi-metal \cite{Kit68,rkky,LLX+09,GCXZ09,KKB17,HKD20,YY18}.

As can be seen in Figs.\ \ref{fig:dberry}, \ref{fig:fit} and \ref{fig:drkky}, differences between topologically nontrivial and trivial case are not very pronounced as concerns the distance dependence. 
This is governed by the finite gap $\Delta E$, which has always been chosen to be the same when comparing both phases.

Qualitatively, the exponential decay of both, the spin-Berry curvature and the RKKY exchange, can be understood easily: 
In the weak-$J$ regime the coupling of two impurity spins results from virtual second-order-in-$J$ processes involving (de-)excitations of electron across the gap $\Delta E$.
This is not only the cause of the exponential distance dependence but also explains the small absolute values of $\Omega / J^{2}$ and $J_{\rm RKKY} / J^{2}$, which do not exceed values of the order of $10^{-2}$.

\subsection{Parameter dependence of the spin-Berry curvature}

The discussion of $M=2$ spin dynamics in Sec.\ \ref{sec:sdyn2} has shown that a value of $\Omega$ close to unity is required for a substantial impact of the geometrical spin torque. 
As we have seen, this is in fact possible in case of finite systems with periodic boundary conditions at small values for $\tau_{2}$ (see Figs.\ \ref{fig:pos1} and \ref{fig:pdtau2}). 
We now focus on large systems again and study the model for small distances between $i_{m}$ and $i_{m'}$ at $\tau_{2}=0.1$, where finite-size effects can be neglected safely.

\begin{figure}[t]
\includegraphics[width=0.9\columnwidth]{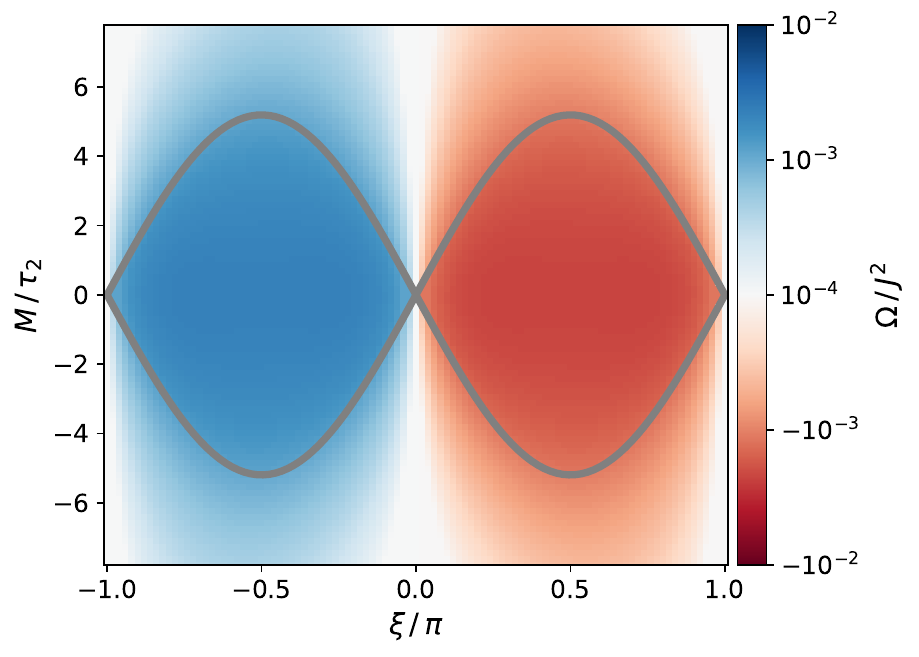}
\caption{
Thick gray lines: phase diagram of the Haldane model in the $\xi$-$M/\tau_{2}$ plane (see Fig.\ \ref{fig:tpd} for comparison).
The color codes the spin-Berry curvature $\Omega / J^{2}$ for next-nearest-neighbor sites $i_{m}$, $i_{m'}$.
Calculations for a system with $27 \times 27$ unit cells with periodic boundary conditions at $\tau_2 = 10^{-1}$.
}
\label{fig:pd1}
\end{figure}

\begin{figure}[b]
\includegraphics[width=0.9\columnwidth]{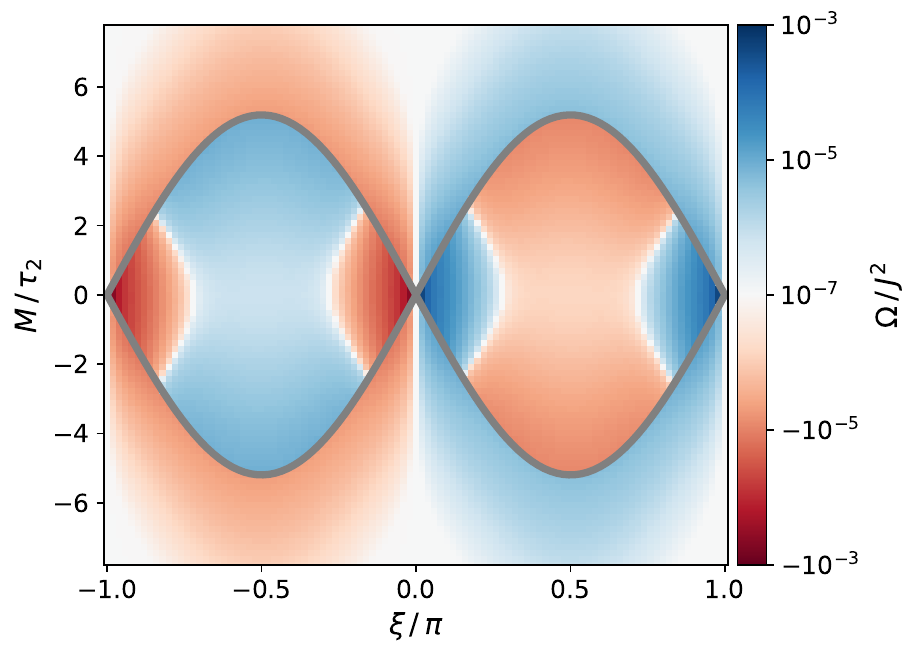}
\caption{
The same as Fig.\ \ref{fig:pd1} but for sites $i_{m}$ and $i_{m'}$ with Euclidean distance $d = 5\sqrt3$ (in units of the nearest-neighbor distance).
}
\label{fig:pd2}
\end{figure}

Fig.\ \ref{fig:pd1} gives an overview over the dependence of the nonzero element $\Omega$ of the spin-Berry curvature on the parameters $M$ and $\xi$ for next-nearest neighbors $i_{m}$, $i_{m'}$.
The boundaries of the topological phase transitions are superimposed in the figure (see thick gray lines).
We find a somewhat larger absolute value of $\Omega$ within a topologically nontrivial phase. 
Furthermore, we note that $\Omega \to -\Omega$ under a sign change of $\xi$. 
As a sign change of phase $\xi$ has the same effect for the Hamiltonian of the Haldane model \refeq{haldane} as a reflection at a mirror symmetry axis of the hexagonal lattice, this observation is easily explained with \refeq{sym}.
Otherwise the  parameter dependence is more or less featureless.
Absolute values for $\Omega$ do not exceed $\sim 10^{-3}$ in the entire parameter regime.

For larger distances between the sites $i_{m}$ and $i_{m'}$, absolute values for $\Omega$ are smaller.
But its parameter dependence can be much more complicated. 
An example is given with Fig.\ \ref{fig:pd2}, which displays results for sites at a distance  $d = \| i_{m} - i_{m'} \| = 5\sqrt3$. 
This must be traced back to the matrix elements in \refeq{rep5}.

In all cases we find that the parameter dependence of $\Omega$ is smooth, including parameter ranges where the model is close to or right at a topological phase transition.
This is worth mentioning since the squared energy denominator in \refeq{rep5}, 
$( \varepsilon_{\rm unocc.}(\ff k') - \varepsilon_{\rm occ.}(\ff k) )^{2}$, suggests that the contributions of wave vectors $\ff k, \ff k'$ at (or close to) the critical point $K$ or $K'$ lead to a diverging or at least large spin-Berry curvature.

\subsection{Spin-Berry curvature close to a topological phase transition}

\begin{figure}[b]
\includegraphics[width=0.9\columnwidth]{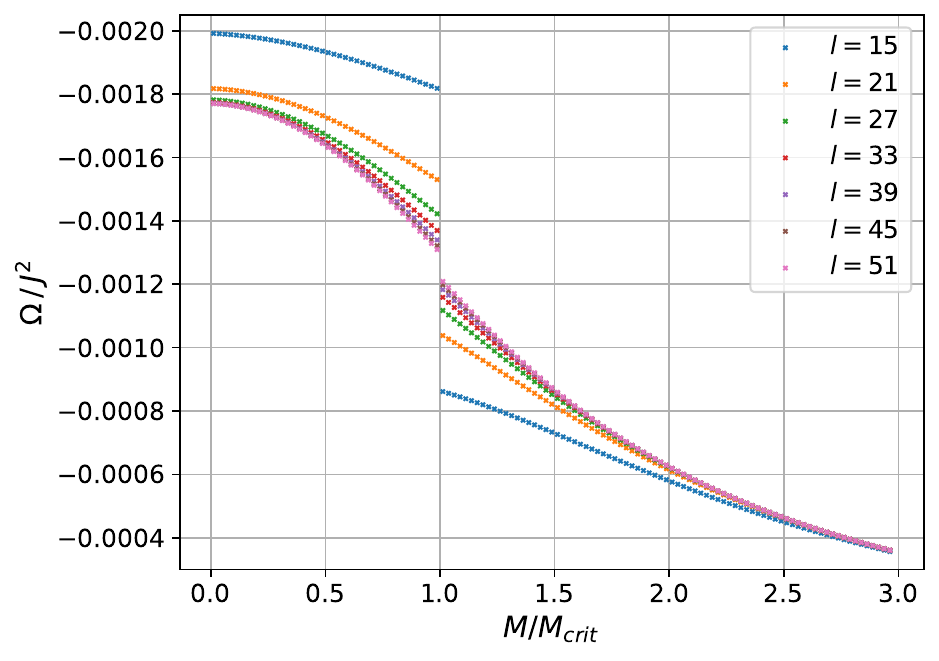}
\caption{
$\Omega$ for next-nearest-neighbor sites as function of $M/M_{\rm crit}$ at $\tau_{2}=0.1$ and $\xi=0.1\pi$, where $M_{\rm crit}$ refers to the critical $M$ value. 
Calculations for system size $L=l\times l$ with various linear extensions $l$ as indicated.
}
\label{fig:oml1}
\end{figure}

Close to a transition, however, a careful analysis of the finite-size effects is necessary.
For systems with a finite number of units cells $L=l\times l$, the spin-Berry curvature is in fact discontinuous at a topological phase transition (actually, the latter is well defined for $L=\infty$ only). 
Fig.\ \ref{fig:oml1} displays results for the $M$ dependence of $\Omega$. 
A finite jump $\Delta \Omega$ at the critical point $M=M_{\rm crit} = 3 \sqrt{3} \tau_{2}  \sin \xi$ is found for various $l$. 
At $l=15$, the relative jump $\Delta \Omega / \Omega$ is considerable. 
With increasing $L$, however, it monotonically but slowly decreases with $l$ and is about an order of magnitude smaller at $l=51$.

\begin{figure}[t]
\includegraphics[width=0.9\columnwidth]{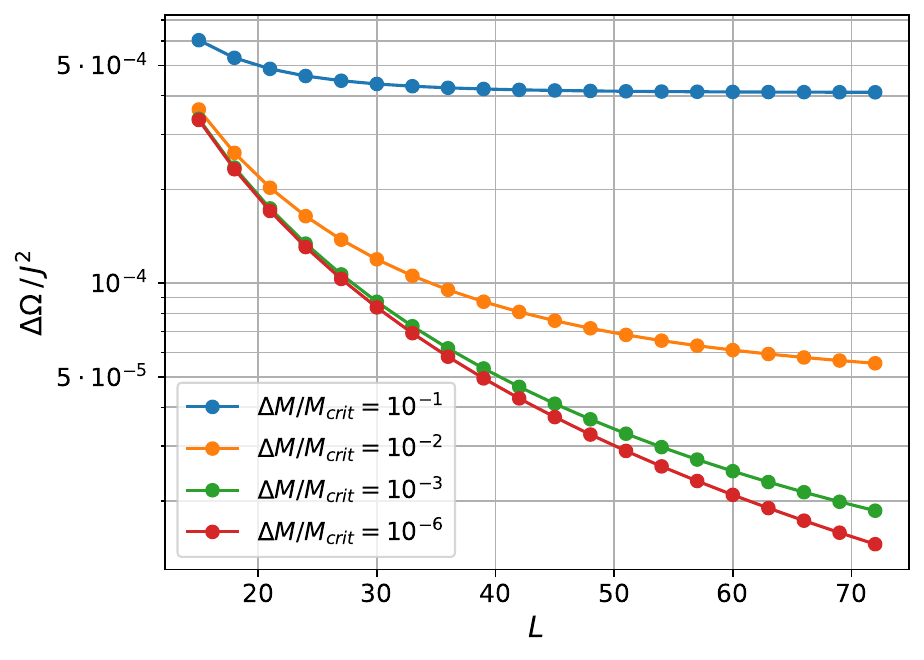}
\caption{
The jump $\Delta \Omega = |\Omega_{\rm top} - \Omega_{\rm triv}|$, obtained by computing the spin-Berry curvature at $M=M_{\rm crit} \pm \Delta M$, as function of $l$ for various $\Delta M / M_{\rm crit}$.
Calculations for next-nearest neighbors, $\tau_2 = 0.1$, $\xi = \pi / 2$.
}
\label{fig:ldep}
\end{figure}

The data are consistent with the proposition that $\Omega$ is continuous at $M=M_{\rm crit}$ in the limit $L\to \infty$.
A numerical proof, however, is difficult to achieve, since besides the $L\to\infty$ limit, the $\Delta M = |M-M_{\rm crit}| \to 0$ limit must be taken simultaneously.
Fig.\ \ref{fig:ldep} displays the jump $\Delta \Omega$ as function of $l$ for various $\Delta M / M_{\rm crit}$. 
For comparatively large $\Delta M / M_{\rm crit}=0.1$, we see that $\Delta \Omega$ decreases but approaches a finite value for $l \to \infty$. 
At smaller relative distances $\Delta M / M_{\rm crit}$ to the transition point, however, this assumed convergence with $l$ eventually becomes invisible for the largest feasible system sizes (see red line).
Nevertheless, continuity of $\Omega$ at a topological phase transition appears highly plausible. 

In fact, an analytical argument can be given:
Parametrically close to a topological phase transition and in the vicinity of the band-closure points $K$ or $K'$ in the first Brillouin zone, the band structure of the Haldane model takes the form of a relativistic Dirac theory \cite{Hal88,Ber13}. 
If $\ff \kappa=(\Delta k_{x},\Delta k_{y})$ and $\ff \kappa'=(\Delta k'_{x},\Delta k'_{y})$ denote the wave vectors {\em relative} to $K$ or $K'$, i.e., if $\ff \kappa,\ff \kappa'=0$ refer to a band-closure point, we have
\be
\varepsilon_{\nu}(\ff \kappa) = \varepsilon_{\pm}(\ff \kappa ) \propto \pm \sqrt{\kappa_{x}^{2} + \kappa_{y}^{2}+m^{2}} \, .
\labeq{dirac}
\ee
The ``mass'' $m$ is linearly related to the insulating gap: $m\propto \Delta$. 
We can analytically check for a possible divergence of the spin-Berry curvature in the thermodynamical limit $L\to \infty$ on approaching the phase transition via $m\to0$ with help of \refeq{rep5}.
To this end, we compute the contribution $I_{\rm bulk}$ of wave vectors in a sufficiently small ball $B$ with radius $\Lambda$ around $\ff \kappa=0$. 
Up to a constant factor, we find
\be
  I_{\rm bulk} \propto \int_{B} d^{2}\kappa \int_{B} d^{2}\kappa' \: \frac{1}{\big(\sqrt{\ff \kappa^{2} + m^{2}} + \sqrt{{{\ff \kappa}'}{}^{2} + m^{2}}\big)^{2}} \: ,  
\ee
and the spin-Berry curvature is continuous at $m=0$, if $\lim_{m\to 0} I_{\rm bulk}$ exists.

Note that to realize the thermodynamical limit, the wave-vector sums in \refeq{rep5} have been replaced by integrations. 
Furthermore, $\nu=-$ (occupied) and $\nu'=+$ (unoccupied), see \refeq{dirac}. 
The $\ff \kappa$ dependence of the matrix elements $S_{\ff Ir,\ff \kappa\nu}$ in the numerator in \refeq{rep5} can be disregarded for small $\Lambda$: 
The first factor of each matrix element is a Fourier factor with a smooth $\ff \kappa$ dependence, see \refeq{comb}. 
For each $\nu$, the second factor is given by the two-component eigenstate of the Dirac Hamiltonian \cite{Hal88,Ber13}
\be
H_{\rm D} \propto \kappa_{y} \sigma_{x} - \kappa_{x} \sigma_{y} + m \sigma_{z}
\: ,
\ee
from which the dispersion \refeq{dirac} is derived. 
Rewriting the two-dimensional $\ff \kappa$ integration (and analogously for the $\ff \kappa'$ integration) with the help of polar coordinates $(\kappa, \varphi)$, i.e., $\int d^2{\kappa}=\int d\kappa \kappa \int d\varphi$, the $\ff \kappa$-dependent part of this factor is of the form $e^{\pm i \varphi}$ and thus cannot lead to a divergence.

The remaining two-dimensional integral
\be
  I_{\rm bulk} = \int_{0}^{\Lambda}\int_{0}^{\Lambda} d\kappa d\kappa' \: \frac{\kappa \kappa'}{\big(\sqrt{\kappa^{2} + m^{2}} + \sqrt{{\kappa'}^{2} + m^{2}}\big)^{2}} \: 
\labeq{divbulk}
\ee
can be computed analytically and turns out to stay finite in the limit $m=0$ with an additive lowest-order correction of the form $- m^{2} \ln m \to 0$ for $m\to 0$.
This implies that the spin-Berry curvature, at a topological phase transition, is a continuous function of the model parameters.

\subsection{Strong exchange coupling}
\label{sec:strong}

The representation \refeq{rep5} for the spin-Berry curvature holds in case of weak exchange coupling $J$. 
For coupling strengths beyond the perturbative regime, we must resort to \refeq{rep2}, where the matrix elements are defined with eigenstates carrying a nonvanishing dependence on the spin configuration.
The numerical evaluation can be performed as described by the text below \refeq{rep5}. 

\begin{figure}[b]
\includegraphics[width=0.99\columnwidth]{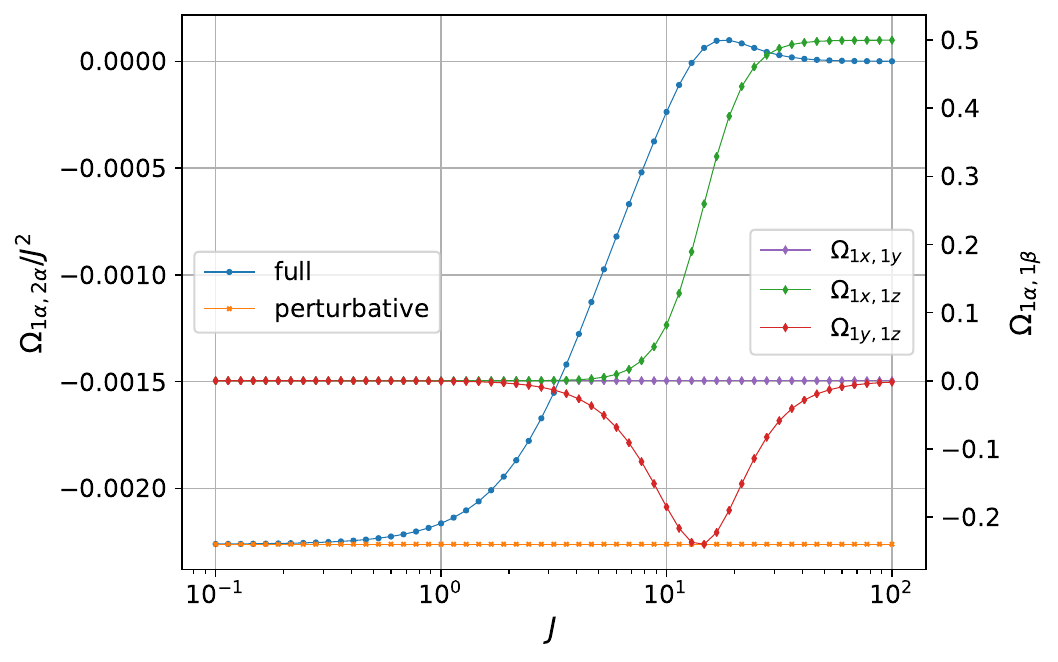}
\caption{
Spin-Berry curvature as function of $J$ for an $L=60 \times 60$ lattice at $M = 0$, $\tau_2 = 0.1$, $\xi = 0.25 \pi$, and for a fixed classical spin configuration
$\ff S_m = (0,1,0)$, $\ff S_{m'} = (-1,0,0)$, where $i_{m}$ and $i_{m'}$ are next-nearest neighbors.
Blue data (see left scale):
Off-site $m,m'$ element $\Omega = \Omega_{m\alpha,m'\alpha'}$ (and $\alpha=x$, $\alpha'=y$), divided by $J^{2}$, to be compared with the corresponding perturbative result (orange data). 
Violet, green, red data (see right scale):
On-site elements $\Omega_{m\alpha,m\alpha'}$ at the position $i_{m}$.
}
\label{fig:strong}
\end{figure}

Fig.\ \ref{fig:strong} displays the spin-Berry curvature as a function of $J$. 
Comparing the results obtained from the full theory (blue data, left scale) with those of the perturbative-in-$J$ approach (orange), we see that perturbation theory applies to coupling strengths $J\lesssim 0.1$ and is still a good approximation up to $J \lesssim 1$.

For even stronger $J$ beyond the perturbative regime, the structure of the spin-Berry-curvature tensor $\Omega_{m\alpha,m'\alpha'}$ changes qualitatively. 
In the $J\to \infty$ limit, only on-site elements $\Omega_{m\alpha, m\alpha'}$ are nonzero, and the blue curve (left scale) approaches zero in this limit.
These on-site elements are given by the spin-Berry curvature of the effective two-spin model $H_{\rm 2-spin} = J \ff s_{i_{m}} \ff S_{m}$ at $i_{m}$, where $\ff s_{i_{m}}$ can be treated as an $s=1/2$ quantum spin. 
The two-spin model is easily solved analytically \cite{SP17}, and $\ff \Omega_{m} \equiv (1/2) \sum_{\alpha\alpha'\alpha''} \varepsilon_{\alpha\alpha'\alpha''} \Omega_{m\alpha,m\alpha'} \ff e_{\alpha''}$ is given by the ``magnetic-monopole field''
\be
  \ff \Omega_{m} = - \frac12 \frac{\ff S_{m}}{|\ff S_{m}|^{3}} = -\frac12 \ff S_{m} 
  \: ,
\ee
since $|\ff S_{m}|=1$ and $J>0$. 
As $\ff S_{m} = (0,1,0)$ has been assumed to point in $y$ direction, we find that only $\Omega_{mx,mz} = - \Omega_{m,y} = + 1/2$ remains nonzero (green data, right scale) for $J\to \infty$.
On the contrary, the perturbative approach, see \refeq{omega}, yields $\Omega_{m \alpha, m' \alpha'} = \Omega_{mm'} \delta_{\alpha\alpha'} = 0$ for the on-site elements $m=m'$ due to the antisymmetry of the tensor.

In the intermediate-$J$ regime, see $J \approx 10$ in Fig.\ \ref{fig:strong}, there is still a finite off-site element (blue, left scale). 
The on-site element $\Omega_{mx,mz} = - \Omega_{m,y}$ (green, right scale) is still far from its asymptotic value for $J \to \infty$, while
$\Omega_{my,mz} = \Omega_{m,x}$ has a finite negative value due to the proximity of $\ff S_{m'}$ pointing into $-x$ direction.
Finally, for the chosen classical spin configuration, $\Omega_{mx,my} = \Omega_{m,z} = 0$ in the entire $J$ range. 

The data in the intermediate-coupling regime ($J \sim \ca O(10)$) demonstrate that the elements of the spin-Berry curvature tensor may well assume values of the order of one. 
They appear to be limited, however, by their $J\to \infty$ values $|\Omega_{m,\alpha}| \le 1/2$.
Nevertheless, one would have a strongly anomalous spin dynamics with it.

\subsection{Results for a ribbon geometry}

We have seen that the spin-Berry curvature stays finite at a topological phase transition, since the gap merely closes at a single critical point, $K$ or $K'$ in the two-dimensional Brillouin zone. 
This is a too weak singularity to have a significant impact on the $\ff k, \ff k'$ sums in \refeq{rep5}.
For the model on a one-dimensional lattice, however, this is qualitatively different.

Here, we consider the Haldane model on a one-dimensional ribbon with a large number of unit cells $L_{x}$ and periodic boundary conditions in the $x$ direction, and with a finite number of unit cells $L_{y}$ and open boundary conditions in the $y$ direction, so that the ribbon is bounded by zigzag edges in the $y$ direction.
Equation (\ref{eq:rep5}) applies accordingly for one-dimensional wave vectors $k_{x},k_{x}'$ and for $I,I'=1,...,L_{y}$. 

\begin{figure}[t]
\includegraphics[width=0.9\columnwidth]{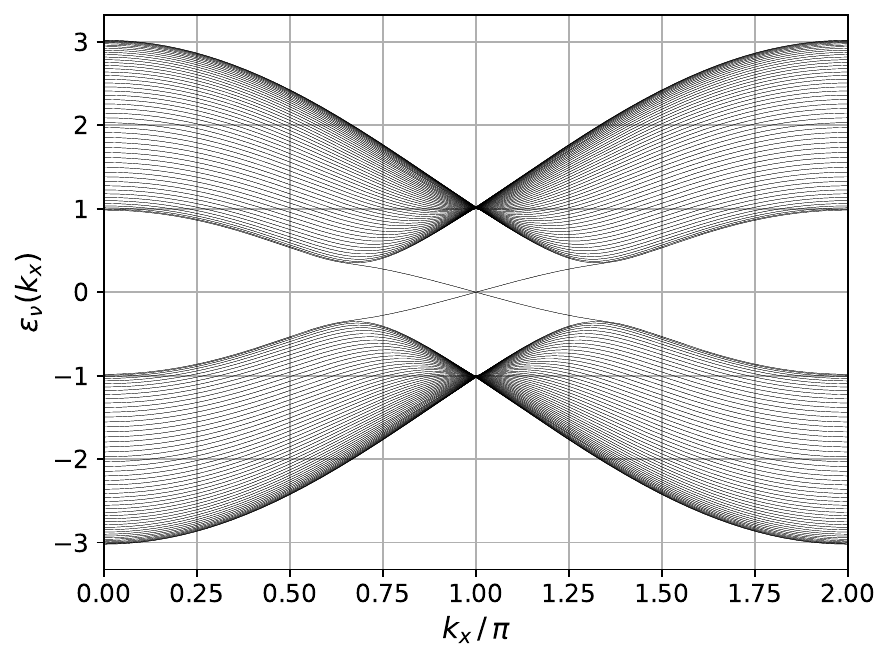}
\caption{
Band structure of a Haldane ribbon with zigzag edges.
Calculations for a ribbon of finite thickness with $L_{y}=50$ unit cells in $y$ direction.
Periodic boundary conditiosn are assumed along the $x$ direction. 
Parameters: $\xi = \pi/2$,  $\tau_2 = 0.1$ and $M=0$. 
}
\label{fig:bs}
\end{figure}

In a topologically nontrivial phase with $C=\pm 1$ and disregarding the trivial spin degeneracy, the bulk-boundary correspondence principle requires the existence of two gapless chiral eigenstates of $\hat{H}_{\rm qu}$ exponentially localized at the opposite edges.
For finite but large $L_{y}$, the band dispersions $\varepsilon_{\nu}(k_{x})$ for $\nu = 1, ..., 2L_{y}$ form two quasi-continua of bulk states in the one-dimensional Brillouin zone separated by the bulk band gap $\Delta E$. 
Within the bulk band gap, the edge-state dispersions $\varepsilon_{1}(k_{x})$ and $\varepsilon_{2}(k_{x})$ take the form of an avoided crossing at low energies. 
With $\kappa \equiv \Delta k_{x}$ we have
\be
\varepsilon_{1,2}(\kappa) \propto \pm \sqrt{\kappa^{2} + m^{2}} \: , 
\labeq{1d}
\ee
with a gap $\propto 2m$ that is exponentially small in $L_{y}$ for large $L_{y}$. 
For $L_{y}\to \infty$ the energy spectrum is gapless ($m=0$), and at low excitation energies the edge-state dispersions are linear with positive and negative slope, respectively:
\be
\varepsilon_{1}(\kappa) \propto \kappa \:  , \quad \varepsilon_{2}(\kappa) \propto -\kappa \: .
\labeq{lin}
\ee
As an example, Fig.\ \ref{fig:bs} shows the ribbon band structure for model parameters, where the spectrum is particle-hole symmetric. 
Fixing the Fermi energy at zero, $\varepsilon_{\rm F}=0$, we have a half-filled system.

Addressing the weak-$J$ limit, we use Eq.\ (\ref{eq:rep5}) to compute the contribution $I_{\rm rib}$ to the spin-Berry curvature, for sites $i_{m}, i_{m'}$ at one of the edges and for $L_{x}, L_{y} \to \infty$, due to relative wave vectors $\kappa$ in a sufficiently small range around $\kappa=0$. 
This is given by $\lim_{\varepsilon \to 0}  I_{\rm rib}(\varepsilon)$ for positive $\varepsilon$, where
\be
  I_{\rm rib}(\varepsilon) 
  =
  \int_{\varepsilon}^{\Lambda} d\kappa \int_{\varepsilon}^{\Lambda} d\kappa' \: \frac{1}{(\kappa+\kappa')^{2}} \: .
\ee
It is straightforward to see that the integral diverges for $\varepsilon \to 0$ as $I_{\rm rib}(\varepsilon) \sim - \ln \varepsilon$, contrary to the two-dimensional case, see \refeq{divbulk}.
We conclude that the spin-Berry curvature is weakly (logarithmically) divergent in the topologically non-trivial phase due to the presence of edge modes.

There are two sources for a finite gap regularizing the spin-Berry curvature. 
First, for a ribbon with a large but finite $L_{x} < \infty$, the $k_{x}$-space discretization, $\delta k_{x} = 2\pi/L_{x}$, will regularize the divergence, such that
$\Omega \sim \ln L_{x}$. 
Second, for $L_{x}\to \infty$, but for a finite ribbon width $L_{y}$, a gap $\propto m < \Delta E$ is produced by overlapping edge wave functions, see \refeq{1d}. 
In this case, $\Omega \sim -\ln m$. 

\begin{figure}[t]
\includegraphics[width=0.9\columnwidth]{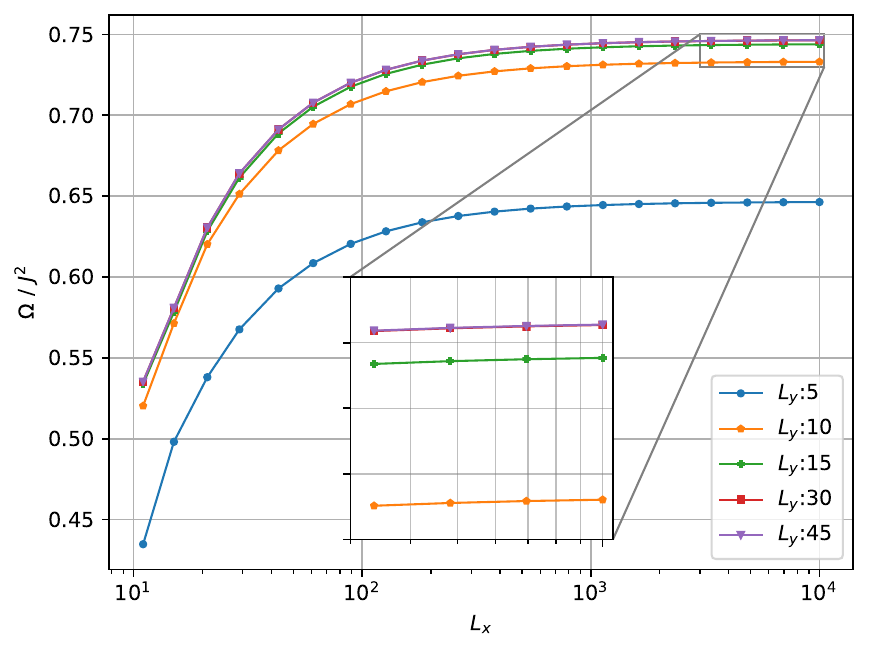}
\caption{
$\Omega$ as function of nano-ribbon length $L_x$ (with periodic boundary conditions in $x$ direction) for various ribbon widths $L_y$ as indicated. 
Note the logarithmic scale for $L_{x}$.
The classical impurity spins are located at next-nearest-neighbor sites at the same zigzag edge. 
Calculations for the particle-hole symmetric case with $M=0$ and with Fermi energy $\varepsilon_{\rm F}=0$. 
Further parameters: $\tau_2 = 0.1$, $\xi = 0.1 \pi$.
}
\label{fig:ribbon}
\end{figure}

Fig.\ \ref{fig:ribbon} displays the nonzero element $\Omega$ of the spin-Berry curvature for two next-nearest-neighbor sites at the same edge as function of $L_{x}$ and for various finite $L_{y}$.
For fixed $L_{x}$ and with increasing $L_{y}$, the overlap between the edge states localized at different edges and thus the gap parameter $m$ decrease exponentially fast with $L_{y}$, such that one expects that the gap due to $k$-space discretization $\delta k_{x} = 2\pi/L_{x}$ quickly becomes the dominating factor.
In fact, for any $L_{x}$, the spin-Berry curvature $\Omega$ as a function of $L_{y}$ quickly converges to a finite value. 
On the scale of the plot, there is no visible difference between the results for $L_{y} = 30$ and $L_{y} = 45$ (see also the red and violet data in the inset).
For the largest $L_{y}$, one notes that $\Omega$ still increases with $L_{x}$ at values of the order of $L_{x} = \ca O(10^{4})$, see inset. 
The dependence of $\Omega$ on $\log L_{x}$ is close to linear, as expected, but with a small slope.

Importantly, the absolute values of $\Omega/J^{2}$ in the wide $L_{x}$-$L_{y}$ range considered are of the order of unity. 
According to \refeq{singu} and if $J$ is of the order of unity as well, this is just the range, where the strongest effects of the geometrical spin torque can be expected. 
We conclude that, compared with the situation of the two-dimensional bulk system close to a topological phase transition, the phase-space reduction in the one-dimensional case is very favorable for a large spin-Berry curvature.

\section{Summary and outlook}
\label{sec:con}

The typical time scale of the dynamics of local magnetic moments, exchange coupled to an electron system, is controlled by the strength of the exchange coupling $J$. 
Since this is usually one or even several orders of magnitudes smaller than the characteristic energy scales of the electron system, the local-moment dynamics is slow as compared to the characteristic electronic time scale.
In particular, if $J\ll \Delta E$, where $\Delta E$ is size of the gap of an insulating electron system, the adiabatic theorem applies, and the ground state of the electron system at an instant of time $t$ is well approximated by the instantaneous ground state corresponding to the configuration of the local moments at time $t$.
Modelling the magnetic moments as classical spins of fixed length, the spin dynamics is described by an effective low-energy classical theory, which is much simpler than the full coupled semiclassical electron-spin dynamics, since it involves ground-state quantities of the electron system only. 

We have developed this effective spin-only theory within a Lagrange formalism. 
Besides the torque resulting from well-known indirect magnetic exchange, there is is an additional spin torque that is given in terms of the spin-Berry curvature. 
This geometrical spin torque is analogous to the geometrical force discussed in molecular dynamics and is nonzero in case of electron systems with broken time-reversal symmetry. 
As we have demonstrated explicitly for the simple case of two classical spins, the emergent dynamics is highly unconventional and differs from the dynamics of any spin-only Hamiltonian.
Future studies may address this non-Hamiltonian spin dynamics for systems like the semi-classical Kondo-{\em lattice} model with a large number of spins.
 
For our present study, which aims at an in-principle demonstration, we have considered the (spinful) Haldane model as a prototypical system, where time-reversal symmetry is broken explicitly.
This choice has the additional benefit that the spin-Berry curvature can be compared with $\ff k$-space Berry curvature, which plays a central role in topological band theory. 
There is in fact a (weak) relation between both, as their Lehmann-type representations involve exactly the same energy eigenstates, albeit different operators must be considered in the matrix elements. 
This explains the sensitivity to the topological phase that has been observed in the study of the spatial structure of the spin-Berry curvature tensor, particularly at the  zigzag edges of the lattice.
For forthcoming work, it will be interesting to consider other condensed-matter systems with broken time reversal symmetry, such as systems with spontaneous magnetic order. 

It has been shown that the spin-Berry curvature tensor, in the weak-$J$ limit, is closely related, namely by the frequency derivative at  $\omega=0$, to the magnetic susceptibility.
This insight opens the possibility to make close contact to well-known properties of the indirect magnetic exchange, i.e., to RKKY theory and to its pendant for insulators and semimetals, e.g., as concerns their distance dependence.

We have systematically studied the magnitude of the spin-Berry curvature. 
As the example for two classical spins has shown, its effects, namely an overall renormalization of the indirect magnetic exchange and an additional coupling between the spins, are most pronounced, if this is of the order of unity. 
Apart from the strong-$J$ regime, however, this is not easily achieved in case of the Haldane model. 
The main reason is that virtual second-order-in-$J$ processes are exponentially damped by the finite gap size $\Delta E$, similar to the exponential $\Delta E$ dependence of the indirect magnetic exchange. 
This suggests that strong effects should be expected for systems in close parametrical distance to a topological transition with a band closure.
However, we could argue that the spin-Berry curvature is finite and continuous at a topological transition of the infinite bulk system. 
On the contrary, an arbitrarily large curvature can be observed for finite systems (with periodic boundaries). 
This underpins the view that a phase-space mechanism is at work:
A large spin-Berry curvature of a condensed-matter system in the thermodynamical limit requires parametric vicinity to a band closure on a $D-1$-dimensional submanifold of the $D$-dimensional Brillouin zone.
In fact, we could observe a logarithmically diverging spin-Berry curvature for the Haldane model at the $D=1$ zigzag edge, where a ($D=0$ dimensional) gap closure  is enforced via the bulk-boundary correspondence.
This insight will be useful for forthcoming studies. 

\acknowledgments
This work was supported by the Deutsche Forschungsgemeinschaft (DFG, German Research Foundation) 
through the SFB 925 (project B5), project ID 170620586, and through the research unit QUAST, FOR 5249 (project P8), project ID 449872909.

\appendix

%


\end{document}